\pdfoutput=1 
\documentclass[%
 aip,
 amsmath,amssymb,
 reprint,%
]{revtex4-1}

\usepackage{graphicx}%
\usepackage{dcolumn}%
\usepackage{siunitx}
\usepackage{bm}%

\usepackage[utf8]{inputenc}
\usepackage[T1]{fontenc}
\usepackage{mathptmx}
\usepackage{etoolbox}
\usepackage{caption}
\usepackage{subcaption}
\usepackage[normalem]{ulem}
\usepackage{multirow}
\usepackage{color}

\usepackage[dvipsnames]{xcolor}
\usepackage[version=3]{mhchem}

\usepackage{tikz}
\usepackage{xparse}
\usepackage{etoolbox}

\newbool{markedup}
\boolfalse{markedup}

\NewDocumentCommand{\includegraphicscorrection}{O{} m}{%
  \ifbool{markedup}{%
    \begin{tikzpicture}[baseline=(image.base)]
      \node[inner sep=0pt] (image) {\includegraphics[#1]{#2}};
      \draw[RoyalBlue, line width=1pt] (image.south west) rectangle (image.north east);
    \end{tikzpicture}%
  }{%
    \includegraphics[#1]{#2}%
  }%
}

\makeatletter
\def\@email#1#2{%
 \endgroup
 \patchcmd{\titleblock@produce}
  {\frontmatter@RRAPformat}
  {\frontmatter@RRAPformat{\produce@RRAP{*#1\href{mailto:#2}{#2}}}\frontmatter@RRAPformat}
  {}{}
}%
\makeatother

\newcommand{\correctioncolor}{%
  \ifbool{markedup}{%
    RoyalBlue%
  }{%
    black
  }%
}

\ifbool{markedup}{%
    \newcommand{\correction}[1]{{\noindent\color{\correctioncolor} #1}}
    \newcommand{\cc}{\color{\correctioncolor}}
}{
    \newcommand{\correction}[1]{{#1}}
    \newcommand{\cc}{}
}

\begin{document}

\title[Learning a transferable kinetic energy functional]{KineticNet: Deep learning a transferable kinetic energy functional\\ for orbital-free density functional theory}
\author{R.~Remme}
\author{T.~Kaczun}
\author{M.~Scheurer}
\author{A.~Dreuw}
\author{F.~A.~Hamprecht}%
 \email{roman.remme@iwr.uni-heidelberg.de}
\affiliation{ 
IWR, Heidelberg University \\
Im Neuenheimer Feld 205, 69120 Heidelberg, Baden-Württemberg, Germany %
}

\date{12 September 2023}

\begin{abstract}
Orbital-free density functional theory (OF-DFT) holds the promise to compute ground state molecular properties at minimal cost. However, it has been held back by our inability to compute the kinetic energy as a functional of the electron density only. 
We here set out to learn the kinetic energy functional from ground truth provided by the more expensive Kohn-Sham density functional theory. 
Such learning is confronted with two key challenges: Giving the model sufficient expressivity and spatial context while limiting the memory footprint to afford computations on a GPU; and creating a sufficiently broad distribution of training data to enable iterative density optimization even when starting from a poor initial guess. 
In response, we introduce KineticNet, an equivariant deep neural network architecture based on point convolutions adapted to the prediction of quantities on molecular quadrature grids.
Important contributions include convolution filters with sufficient spatial resolution in the vicinity of the nuclear cusp, an atom-centric sparse but expressive architecture that relays information across multiple bond lengths; and a new strategy to generate varied training data by finding ground state densities in the face of perturbations by a random external potential. 
KineticNet achieves, for the first time, chemical accuracy of the learned functionals across input densities and geometries of tiny molecules. 
For two electron systems, we additionally demonstrate OF-DFT density optimization with chemical accuracy.
\end{abstract}

\maketitle

\onecolumngrid
\noindent
\fbox{
\parbox{\dimexpr\textwidth-2\fboxsep-2\fboxrule}{
This article may be downloaded for personal use only. Any other use requires prior permission of the author and AIP Publishing. This article appeared in \textit{The Journal of Chemical Physics} 159, 144113 (2023) and may be found at \url{https://doi.org/10.1063/5.0158275}.
}
}
\vspace{10pt}
\twocolumngrid

\section{Introduction}
Kohn-Sham density functional theory (KS-DFT) has become the workhorse of quantum chemistry thanks to its appealing trade-off of computational cost vs.~accuracy of molecular property predictions. Even so, its use of orbitals and resulting cubic scaling with system size precludes its application to larger systems with thousands of atoms that are needed to faithfully model, e.g., macromolecules in solution. 
The main reason that KS-DFT needs orbitals in the first place is that, despite decades of theoretical work, a concrete recipe to accurately compute the non-interacting kinetic energy $T_s$\cite{parr1995density} from the electron density has remained elusive; whereas it can be computed from Kohn-Sham orbitals $\phi_i$ via $T_s=\int t_s(\mathbf{r}) d^3\mathbf{r}$ with a kinetic energy density 
\begin{eqnarray}
    t_s(\mathbf{r}) &= \frac{1}{2} \sum_{i=1}^N |\nabla \phi_i(\mathbf{r})|^2\,. \label{eq:ked}
\end{eqnarray}
Yet, the mesmerizing promise of the Hohenberg-Kohn theorems is that it suffices to solve a single integrodifferential equation for the density $\rho(\mathbf{r})$ to find the ground state of a system, provided we find a concrete means to compute $T_s$ and the kinetic potential (its functional derivative with respect to the electron density $\frac{\delta T_s}{\delta \rho}$) as a functional of the electron density only. 

Extensive theoretical and experimental work has shown that the kinetic energy density is not merely local or ``semi-local", i.e., $t_s(\mathbf{r})$ is not a function of the electron density $\rho(\mathbf{r})$ and its spatial derivatives only. On the other hand, aromatic systems and conductors aside, chemistry exhibits a large degree of locality, suggesting that it should be possible to learn a kinetic energy density functional that generalizes across relevant swathes of chemical space. 

In response, we here propose a deep equivariant neural network architecture to approximate the kinetic energy density~$t_s(\mathbf{r})$. 
Specifically, we make the following contributions: 
\begin{itemize}
\item We propose an equivariant deep architecture ingesting an electron density represented on a quadrature grid along with nuclear locations and charges, and predicting a kinetic energy density on the same grid. 
\item We show how to generate varied electron densities and associated kinetic energy potentials needed to achieve convergence when initiating density optimization far from the ground state. 
\item We demonstrate orbital-free density optimization in systems with two electrons, reproducing bonding with chemical accuracy.
\item We offer machine learned functionals for the kinetic energy density and gradient which yield chemical accuracy in OF-DFT calculations, generalizing over input electron densities, external potential and molecular geometry.
\end{itemize}

\subsubsection*{Related Work}
Machine learning has been used to improve DFT pipelines before. A large number of works\cite{kirkpatrick2021pushing,nagai2020completing,bystrom2022cider,dick2020machine,mardirossian2014omegab97x,schmidt2019machine}
focus on learning an approximation to the exchange correlation (XC) functional, where \citeauthor{kirkpatrick2021pushing}\cite{kirkpatrick2021pushing} recently demonstrated impressive results.
\citeauthor{dick2020machine}\cite{dick2020machine} use an architecture that is similar in some respects to learn the XC functional. However, they move to invariant features early on in their model and do not learn the atomic representations, relying instead on hand-crafted features to encode the atomic environments. They train their model to only predict the total XC energy as a scalar, and compute the XC-potential in a variational manner by back-propagating through their model.
A number of works demonstrate the potential of ML for OF-DFT on one-dimensional data sets, such as \citeauthor{snyder2013orbital}\cite{snyder2013orbital} \citeauthor{meyer2020machine}\cite{meyer2020machine} and \citeauthor{saidaoui2020direct}\cite{saidaoui2020direct}.
The approach of \citeauthor{ghasemi2021artificial}\cite{ghasemi2021artificial} works in 3D, but on single, rotationally symmetric atoms only, effectively reducing the dimensionality to one.
\citeauthor{golub2019kinetic}\cite{golub2019kinetic} take a semi-local approach to the 3D problem as they train a neural network that takes five features reflecting the electron density, its gradient as well as its Laplacian to model the kinetic energy density, which they apply to each grid point individually.
\citeauthor{seino2018semi}\cite{seino2018semi} and \citeauthor{fujinami2020orbital}\cite{fujinami2020orbital} show promising results for learning the kinetic energy and potential for molecules, however their models generalize only over different densities and hence only work for a single molecule with fixed geometry at a time.
\citeauthor{ryczko2022toward}\cite{ryczko2022toward} learn the kinetic functional on a voxel-grid representation that works well for their application to graphene lattices, but is less suited for molecules. They present one of the few approaches with successful density optimization, however only for a learned functional that is trained to mimic the flawed Thomas-Fermi approximation.
The most extensive results including density optimization are presented by \citeauthor{imoto2021order}\cite{imoto2021order}. Like \citeauthor{golub2019kinetic}, they use a simple neural network applied pointwise and learn an enhancement factor to the Thomas-Fermi (TF) functional in a way that guarantees both correct scaling and asymptotic behaviour. They outperform classical approximations, but not by the extent required to reach chemical accuracy.

\section{KineticNet: a deep equivariant architecture}
\begin{figure*}[t]
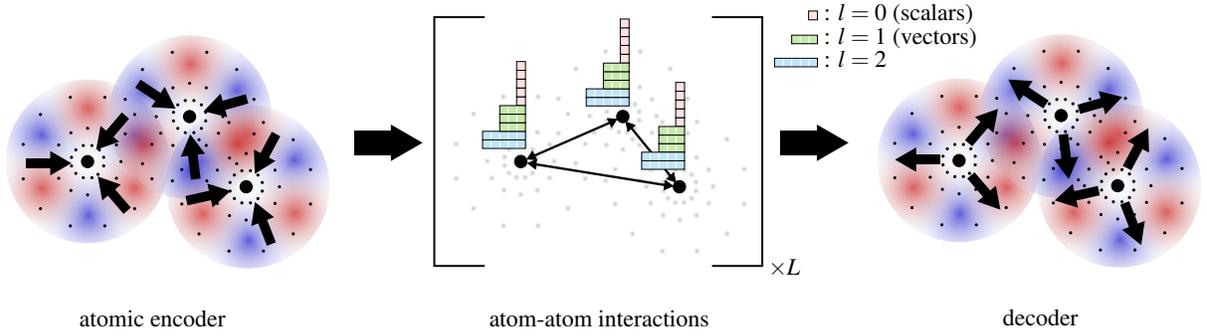
\caption{\label{fig:architecture}The proposed KineticNet architecture is an equivariant deep neural network with three types of layers: First, an atomic encoder relying on point convolutions (eq.~\ref{eq:pointconv}) to summarize the density information on the quadrature grid in terms of tensorial features associated with the nuclei; then a number $L$ of atom-atom interactions layers; and finally a decoding layer making predictions at all grid points.}
\end{figure*}
When developing the architecture for our machine learning model (figure \ref{fig:architecture}), we were guided by a number of physically motivated constraints: Firstly, input and output should be represented on the quadrature grid (consisting of grid points evenly distributed on each of a number of spherical shells arranged around each atomic nucleus), such that it can seamlessly replace existing functional approximations.
Secondly, the model should be equivariant with respect to the group $E(3)$, i.e.\ rotations and translations of the input molecule should not change the predicted kinetic energy, and the predicted kinetic potential should be transformed in accordance with the input.
Finally, the field of view, i.e.\ the spatial extent of the input grid points that influence the output at a given point, should span several bond lengths.
On the other hand, the model should still be local in the sense that for very big molecules, only nearby atoms influence the prediction, thus conceptually allowing for the generalization towards bigger molecules.

We guarantee translational equivariance by only using relative positions in our model, and rotational equivariance by using equivariant convolutions as presented in Tensor Field Networks \cite{thomas2018tensor} and implemented in the e3nn library \cite{geiger2022e3nn}.
This amounts to decomposing convolutional filters $F$ into a radial part $R$ depending on the distance $r=\|\mathbf{r}\|$ and an angular part $Y$, depending on the direction $\hat{\mathbf{r}}=\mathbf{r}/r$. The former is learned and the latter is given by the spherical harmonics (depending on the representation of the in- and output features of the convolution):
\begin{align}\label{eq:pointconv}
F_{cm}^{(l_f, l_i)}(\mathbf{r}) = R_c^{(l_f,l_i)}(r) \, Y_m^{(l_f)}(\hat{\mathbf{r}})
\end{align}
with non-negative integer rotation orders of the input and filter $l_i$ and $l_f$, channel index $c$ and order inside the representation $m\in\{-l_f,..,l_f\}$.
Multiplying the filters with the input features and computing a certain linear combination, using Clebsch-Gordan coefficients as weights, yields equivariant output features of the point convolution. 
We learn separate convolutional filters for each element and use tensorial features up to order $l=4$.

To achieve a sufficient field of view while keeping the computational cost tractable, we use an encoder-decoder structure: In a first atomic encoding layer, we use a point convolution to compute features at every atom of the molecule (and not every input grid point). 
This is followed by a number of atom-atom interaction layers, each of which consists of a point convolution with the atomic nuclei positions as in- and outputs, followed by a nonlinear activation function. 
These layers are computationally cheap and greatly increase the field of view. 
Finally, a decoding layer, again a single point convolution, propagates the information back to the quadrature grid. 
This architecture has a sufficient field of view to capture functional groups and some of their context in molecules, while still being local and allowing for the generalization over molecule compositions. 
The learned atomic encoding layers are one advantage over prior work, as most commonly\cite{dick2020machine,grisafi2018symmetry} handcrafted features are used to encode the local environments of the atoms.
When predicting energy densities, we additionally scale the output with a Superposition of Atomic Densities (SAD) (commonly used as an initial guess in KS-DFT), allowing the model to predict the correct asymptotic behaviour for larger distances from the atoms. 
In particular, the prediction of very small values becomes possible in low-density regions without extremely precise tuning of the parameters of the radial models, which would otherwise be necessary.

As a loss we use a smooth L1-loss, applied to the point-wise difference between the kinetic energy density and potential predictions and the corresponding ground truth on the grid. 
We use an adaptive scale parameter for the transition between the quadratic and linear regimes of the loss, the parameter grows linearly with the target value, but we threshold this value with $10^{-6}\,\text{Ha}/\text{Bohr}^3$ from below, such that the quadratic region can be reached even at grid locations with very small target values (e.g.\ far away from the nuclei).

\begin{figure}
        \centering
        \begin{subfigure}[b]{8cm}
            \centering
            \includegraphics[width=\textwidth]{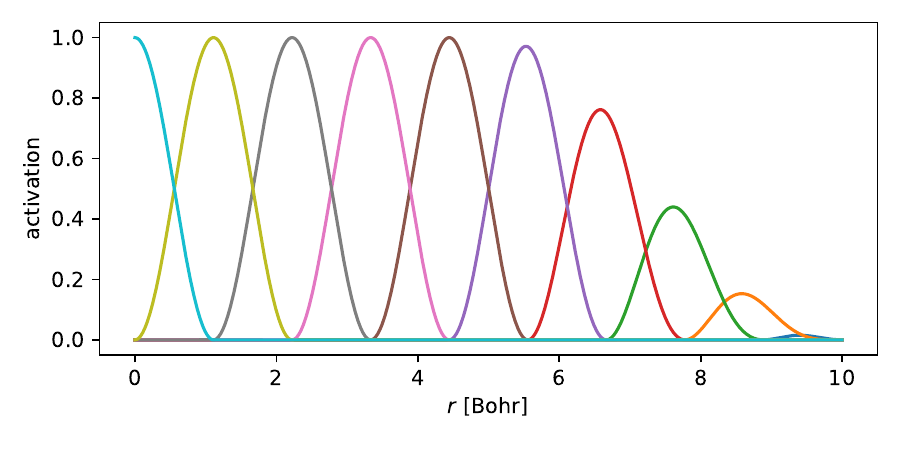}
            \caption[]%
            {{\small radial basis}}    
            \label{fig:untransformed radial model}
        \end{subfigure}
        \vskip\baselineskip
        \begin{subfigure}[b]{8cm}
            \centering 
            \includegraphics[width=\textwidth]{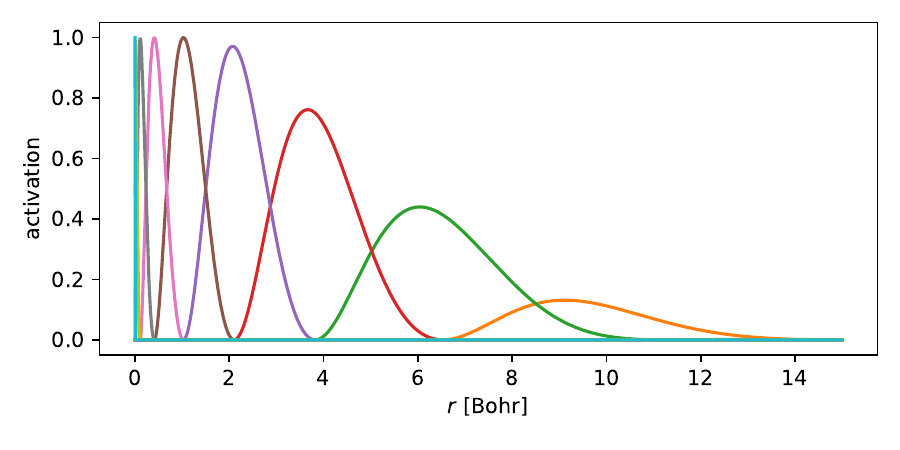}
            \caption[]%
            {{\small transformed radial basis}}    
            \label{fig:transformed radial model}
        \end{subfigure}
        \caption[]
        {\small Schematic of the radial basis to model $R$ in eq.~\ref{eq:pointconv}, with and without transformation to adjust to the Treutler-Alrichs shells, as used in the atomic encoding and decoding layers. We apply a smooth cutoff towards the maximum radius.} 
        \label{fig:radial models}
    \end{figure}
As mentioned above, the parameters of KineticNet lie in the radial models. We parameterize them as \citeauthor{weiler20183d}\cite{weiler20183d}, by a 3-layer fully connected MLP applied to the radius encoded by a set of cosine basis functions. 
For the initial atomic encoding and final decoder layer, we additionally transform the input distances $r$ with the inverse of the Treutler-Ahlrichs\cite{ahrichs1995integration} map $f_{\text{TA}}$ before feeding them to each radial model $R_i$ (where $i$ is a shorthand for indices $c, l_f, l_i$ in eq.~\ref{eq:pointconv}):
\begin{equation}\label{eq:transformed_radial}
    \hat{R}_i(\mathbf{r}) = R_i\left(f_{\text{TA}}^{-1}(\mathbf{r})\right) \quad .
\end{equation}
This effectively changes the radial model to have a distance-dependent spatial resolution (see figure~\ref{fig:radial models}), exactly in correspondence to the spherical shells of the quadrature grid around the atoms.

One feature of our method is that the learning of a spatial filter in terms of absolute distances allows us dealing with varying grid resolutions, i.e.\ spacing of radial shells and angular grids. We do not exploit this explicitly in this work, but it can be useful to speed up the training by first utilizing lower-resolution samples before fine tuning in the high resolution setting, as well as granting the added flexibility of allowing a single model to be deployed at multiple grid resolutions.

\section{Training Data generation \label{sec:DataGeneration}}
Sufficiently large and representative data sets are as decisive for the success of a machine learning approach as the training setup and architecture. 
We use KS-DFT employing the BLYP XC functional\cite{becke1988density,lee1988development} with the cc-pVDZ basis set\cite{dunning1989a, woon1994a} to generate ground truth data for the supervised training of our functionals.
Generating a large number of training samples is easy, but to ensure sufficient variability in the data, we had to employ a new technique that we discuss in this section. 

We use eq.~\ref{eq:ked} to generate ground truth kinetic energy density. Many other definitions exist, in particular any additive constant to $t_s$ that integrates to zero yields the same total kinetic energy. 
That said we choose eq.~\ref{eq:ked} over other formulations for the kinetic energy density as its values lie in a smaller range, which is preferred for machine learning models.
The kinetic potential $\frac{\delta T_s}{\delta \rho}$ can be computed\cite{king2000kinetic} from 
\begin{align}
    \frac{\delta T_s}{\delta \rho} - \mu &= \frac{\sum_i^{N} -\frac{1}{2}\phi_i(\mathbf{r}) \nabla^2 \phi_i(\mathbf{r}) - \epsilon_i \phi_i^2(\mathbf{r})}{\rho(\mathbf{r})} \label{eq:kp}
\end{align}
where $\epsilon_i$ stands for the eigenvalue/orbital energy of the $i$-th KS-Orbital, $\rho$ for the electron density and $\mu$ for the chemical potential, which is assumed equal to the energy of the highest occupied molecular orbital $\epsilon_\text{HOMO}$.\cite{king2000kinetic}

The derivation by \citeauthor{king2000kinetic}\cite{king2000kinetic} equates parts of the Euler and Kohn-Sham equations, suggesting that the equation is only valid for stationary states. 
Yet any OF-DFT algorithm will encounter non-stationary electron densities on its way from the initial guess to the true ground state. As generalization from ground-state densities to these intermediate states cannot be expected, it is crucial to also include 
non-ground state electron densities in the training set.
Such training makes the model sufficiently robust to achieve convergence of the iterative density optimization. This necessity has also been noted by 
\citeauthor{ryczko2022toward}\cite{ryczko2022toward} who observe convergence only for a functional trained to mimic the TF approximation on a varied data set, but not for the functional trained on KS ground truth at ground states only.
In summary, the paradoxical task is to compute the kinetic potential for densities other than the true ground state while at the same time eq.~\ref{eq:kp} holds only for stationary states.
This is where our second contribution lies.

The first Hohenberg-Kohn theorem states that a one to one mapping exists between the external potential and the ground state electron density of a system.\cite{hohenberg_inhomogeneous_1964}
Thus, slightly perturbing the external potential of a molecule will lead to a different electron density as ground state and thereby enable the use of eq.~\ref{eq:kp}. 
Exploiting this observation, we perturb the external potential $v_\text{ext}^\text{mol}(\mathbf{r})$ in KS-DFT by adding a randomly sampled symmetric matrix $\mathbf{M}$ with an appropriately chosen norm to the matrix representation of the external potential in the atomic basis $\left\{ \chi_\nu \right\}$ to generate our training data:
\begin{align}
    \left[ v_\text{ext} \right]_{\mu\nu} &= \langle \chi_\mu | v_\text{ext}^\text{mol} | \chi_\nu \rangle + M_{\mu\nu} \\
    \mathbf{v}_\text{ext} &= \mathbf{v}_\text{ext}^\text{mol} + \mathbf{M}
\end{align}

The pyscf software package\cite{sun2018pyscf, sun2020recent} is used for this purpose as it is efficient and well suited for the integration of ML models trained with python.

\begin{figure*}[t]
\includegraphics[width=0.9\textwidth]{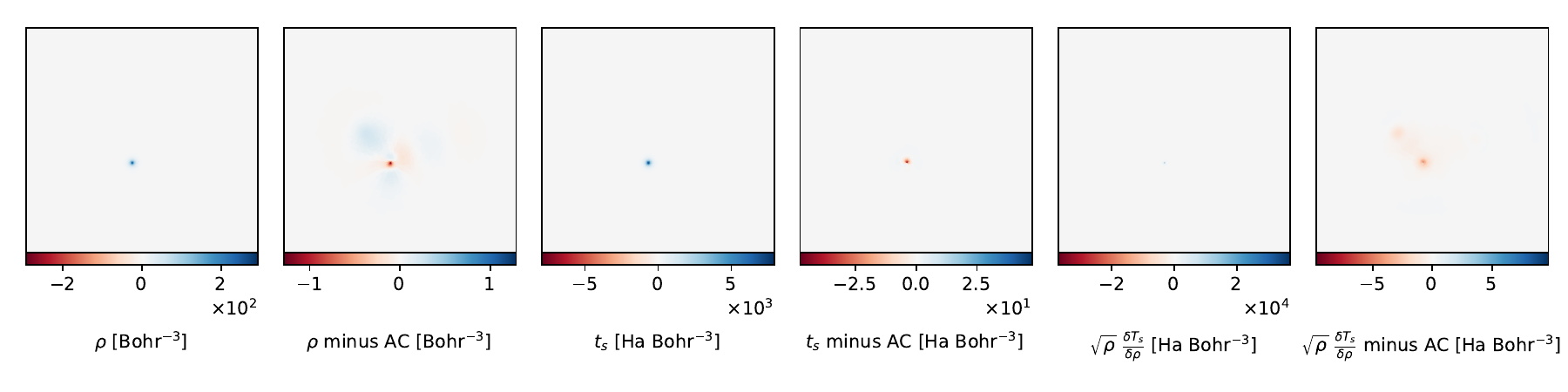}
\caption{\label{fig:atomic_contriubutions_H2O} Effectiveness of subtracting atomic contributions demonstrated for \ce{H2O}. For electron density, kinetic energy density as well as our target for the kinetic potential, subtracting the ACs decreases the value range by at least two orders of magnitude.}
\end{figure*}

For our model, as with most neural networks, it is useful if the inputs and targets have similar scales, and that their values do not vary over many orders of magnitude within a single sample (and between samples). 
Hence, an important detail in the training data generation is how we deal with the cusps at the atoms, of both the input electron density and the output energy density and its potential. 
Here, we take the approach of subtracting spherically symmetric ``Atomic Contributions'' (ACs) for each atom and each of the fields. 
We compute them by applying restricted or restricted open shell KS-DFT to each atom type, and spherically symmetrizing the result, for details see appendix D (in the few cases, in which KS-DFT did not converge, these non-converged solutions still fulfill their purpose). 
This greatly reduces the magnitude of the cusps, see figure \ref{fig:atomic_contriubutions_H2O}. 

Another relevant detail is the choice of target for the kinetic potential: We follow \citeauthor{ryczko2022toward}\cite{ryczko2022toward} and do not directly predict the kinetic potential $\frac{\delta T_s}{\delta \rho}$, but rather $\sqrt{\rho} \frac{\delta T_s}{\delta \rho}$, its product with the square root of the electron density.
They report that this gives a lower training error, and we have two additional reasons to make this choice: 
On the one hand, the denominator in eq.~\ref{eq:kp} leads to numerical problems for small densities, e.g.~far from the atomic nuclei, which are alleviated by taking this product. 
On the other hand, in our OF-DFT calculations, the kinetic potential is multiplied with the square root of the density whenever evaluated, due to our Ansatz (eq.~\ref{eq:of-dft-ansatz}), see section \ref{sec:density optimization} below, hence directly predicting this quantity is natural.

We generate data sets for a number of different atoms and molecules: First, the two-electron systems \ce{He}, \ce{H2} and \ce{H3+}, and furthermore the molecules \ce{HF}, \ce{H2O} as well as two neon atoms in the vicinity of each other as an instance of a non-binding system, which we label as \ce{Ne2}.
We sample the perturbation of the external potential and the molecule geometry independently for each training instance. 
For the linear molecules, we sample the inter-atomic distance uniformly in a range from around 0.4~\textup{\AA} to around 2.0~\textup{\AA}, thus covering both strongly compressed structures as well as nearly dissociated ones. 
For \ce{H3+}, we arrange the nuclei in an equilateral triangle of side length $\sqrt{2}$~\textup{\AA} and perturb the position of each atom by a random vector with a length that is sampled uniformly in $[0, 0.5\text{\textup{\AA}}]$.
Lastly, for water, we apply the following procedure: Each \ce{O-H} bond length is uniformly sampled between 90\% and 110\% of its equilibrium value.
The bond angle is varied by uniformly sampling each \ce{O-H} ``vector'' from a 10° conus.
\correction{Thus, we sample geometries densely around the ground states within arbitrarily defined ranges. In future production level studies, one should carefully design training sets depending on the target application.}

For each system, we generate 8000 training samples, 2000 to use for validation and an additional 1000 for testing.
\correction{

There can be no bond between two atoms if the electron density becomes zero along a surface in between them, and at such zero sets both eq. \ref{eq:kp} for the kinetic potential as well as the von Weizsäcker functional are invalid. Hence, in the two electron case we remove these samples with a zero crossing in the single ``orbital'' $\phi$ (see eq. \ref{eq:of-dft-ansatz} below) which sometimes arise due to the artificial perturbations we add to the external potential when generating the data.
}

\section{Density Optimization}\label{sec:density optimization}
After successfully training these models, the next logical step is to use them ``in the wild'', i.e.~in an actual OF-DFT calculation to compute the ground state of a geometry not seen at training time. 
To this end, we have implemented an OF-DFT solver based on the work of \citeauthor{chan_thomasfermidiracvon_2001}\cite{chan_thomasfermidiracvon_2001} and \citeauthor{ryley_robust_2021}\cite{king2000kinetic} to see if density optimization is possible. 
We use the approach in which the density $\rho$ is represented as the square of a single ``orbital'', or more precisely of a linear combination \correction{$\phi$} of atomic basis functions~$\chi_{\nu}$:
\begin{equation}\label{eq:of-dft-ansatz}
    \rho(\mathbf{r}) = \correction{\phi(\mathbf{r})^2} = \Big(\sum_{\nu} c_{\nu} \chi_{\nu}(\mathbf{r}) \Big)^2 \,.
\end{equation}
The coefficients $c_{\nu}$ are the variables which are optimized to minimize the energy functional while ensuring the correct normalization of the density.
This approach allows the use of well established quantum chemical libraries for the evaluation of many of the integrals.

To achieve this optimization of the total energy w.r.t.~the electron density under the constraint of its normalization to the correct number of electrons $N_e$, a Lagrange multiplier $\mu$ is introduced:
\begin{equation}
\begin{split}
    \mathcal{L} = &T_s[\rho] + \int v_\text{ext}(\mathbf{r}) \rho(\mathbf{r})\, \text{d} \mathbf{r} + J[\rho] + E_\text{xc}[\rho] \\&- \mu \left(N_e - \int\rho(\mathbf{r}) \text{d} \mathbf{r}\right)
\end{split}
\end{equation}
where $T_s$ is the non-interacting kinetic energy, $J$ the classical electron-electron interaction, and $E_\text{xc}$ the exchange correlation energy.
The ground state electron density is then given by the global minimum of this equation.
Therefore, its functional derivative w.r.t.~the electron density gives us the stationarity condition
\begin{align}
    \frac{\delta \mathcal{L}}{\delta \rho } = 0 = \frac{\delta T_s}{\delta \rho} + v_\text{ext}(\mathbf{r}) + \frac{\delta J}{\delta \rho} + \frac{\delta E_\text{xc}}{\delta \rho} - \mu \,.
\end{align}
Introducing the basis expansion, the gradient w.r.t.~the expansion coefficients is then given by
\begin{align}
    \frac{\partial\mathcal{L}}{\partial c_\sigma} &= 2 \langle \chi_\sigma \mid \frac{\delta T_s}{\delta \rho} + v_\text{ext}(\mathbf{r}) + \frac{\delta J}{\delta \rho} + \frac{\delta E_\text{xc}}{\delta \rho} - \mu | \sum_\nu c_\nu \chi_\nu \rangle\,.
\end{align}
Furthermore, the Lagrange multiplier $\mu$, which corresponds to the chemical potential, needs to be optimized:
\begin{align}
    \frac{\partial \mathcal{L}}{\partial \mu} = N_e - \int \rho(\mathbf{r})\, \text{d} \mathbf{r} \,.
\end{align}
 To iteratively solve this constrained optimization problem, we use the SLSQP solver\cite{kraft1988software} as implemented in the scipy package\cite{2020SciPy-NMeth}. 

 For the initial guess in our optimizations we use an adapted version of a SAD guess.
 For this we use the atomic KS-DFT densities and fit OF-DFT density coefficients to it. 
 Those coefficients are then used to construct a guess by simply placing them at the position of the corresponding atomic basis functions. 

\section{Computational experiments}
\begin{figure}
\includegraphics[width=8cm]{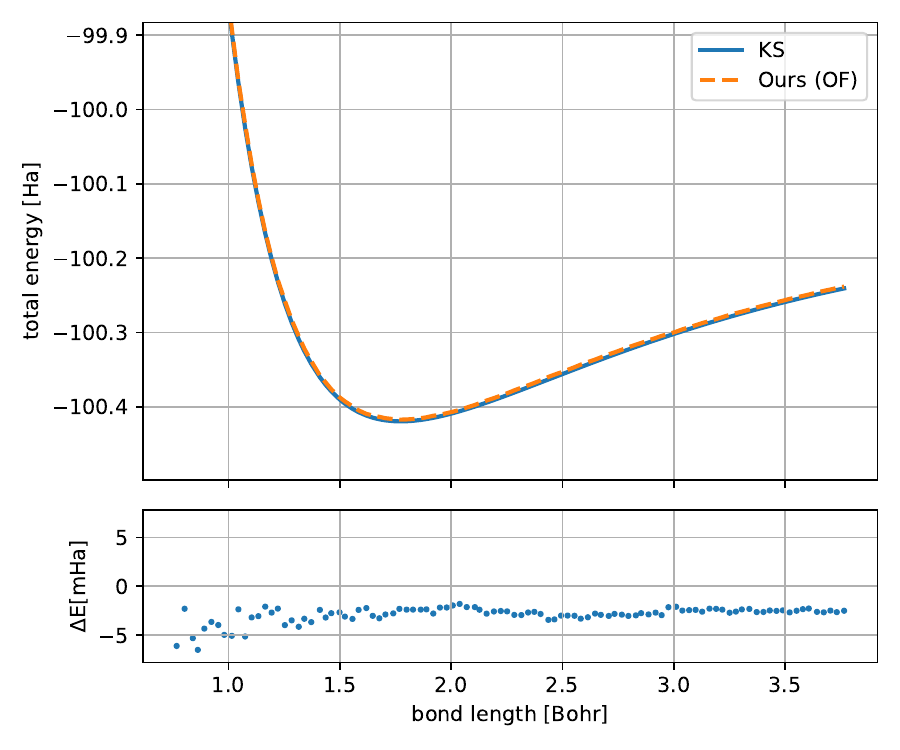}
\caption{\label{fig:hf_dissociation} Total ground state energy of $\ce{HF}$ at different bond lengths, as computed by KS-DFT as well as the prediction of our ML functional on the KS ground-state densities (without orbital-free density optimization).}
\end{figure}

\begin{figure}
\includegraphics[width=8cm]{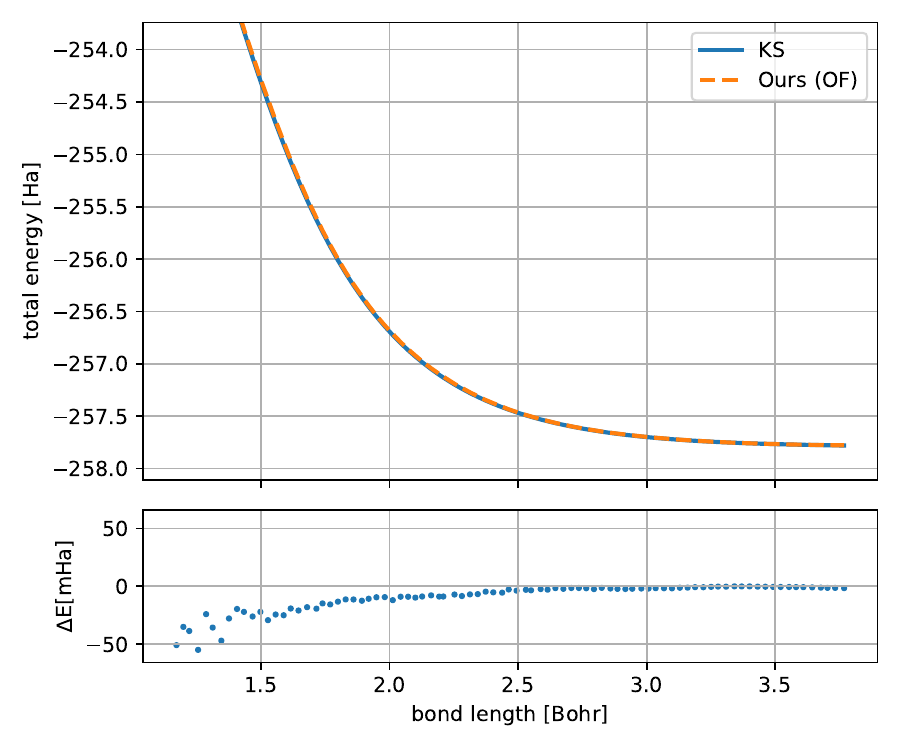}
\caption{\label{fig:ne_dissociation} Total ground state energy of $\ce{Ne2}$ at different ``bond'' lengths, as computed by KS-DFT as well as the prediction of our ML functional on the KS ground-state densities (without orbital-free density optimization).}
\end{figure}

\begin{figure}
\includegraphics[width=8cm]{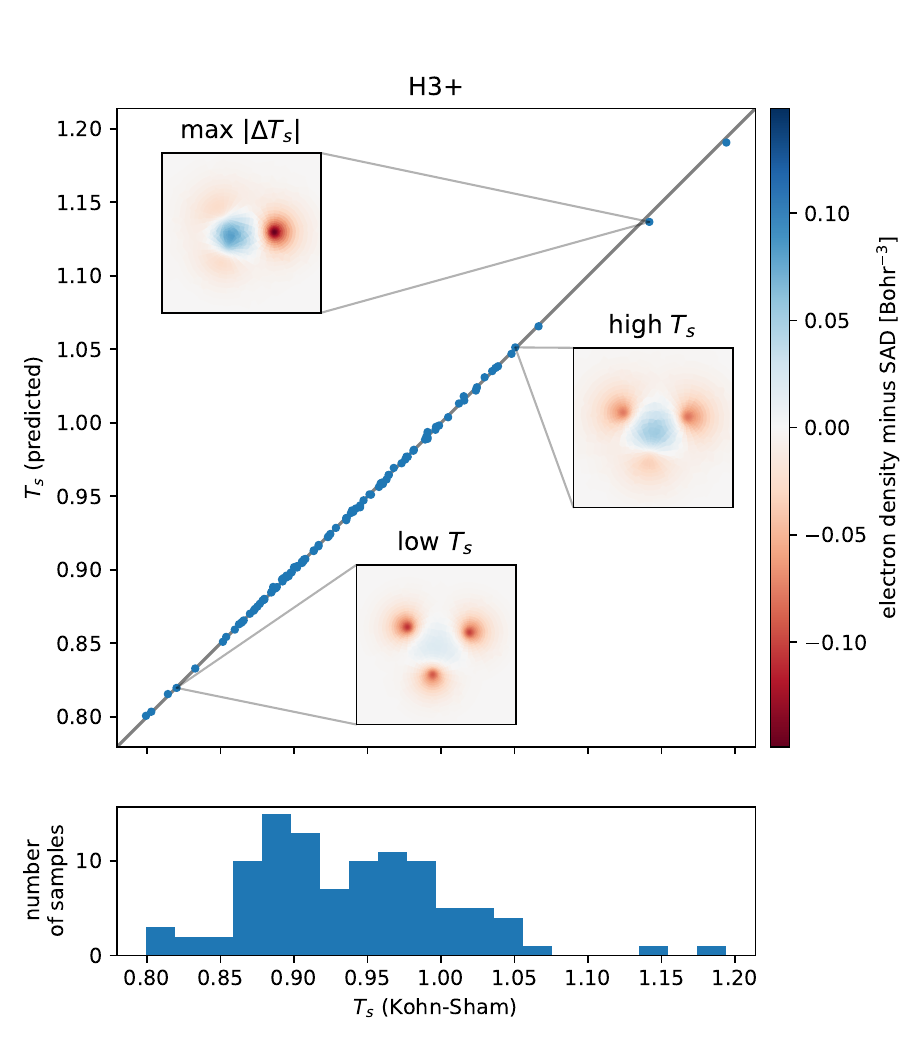}
\caption{Comparison between predicted and KS kinetic energies for \ce{H3+} on 100 samples from the test set, and representative electron densities (minus SAD).\label{fig:scatter_ekin_h3+} }
\end{figure}
\subsection{Training details}
We train, simultaneously but independently, two models: One to predict the kinetic energy density $t_s$ (to be integrated to the kinetic energy $T_s$) and one for the kinetic potential $\frac{\delta T_s}{\delta \rho}$. Each of these models is trained on the union of all data sets.
We train our models using the Adam optimizer\cite{kingma2014adam} using default parameters and a learning rate of $0.01$ that we decay exponentially after $10^5$ training iterations. 
We use a batch size of 64 and train until convergence. 
We observe some amount of overfitting in the sense that the loss is greater during validation than training, however both decrease during the whole training procedure which allows us to simply evaluate the final saved model.

\subsection{Test results}
\begin{figure*}[t]
\includegraphics[width=0.9\textwidth]{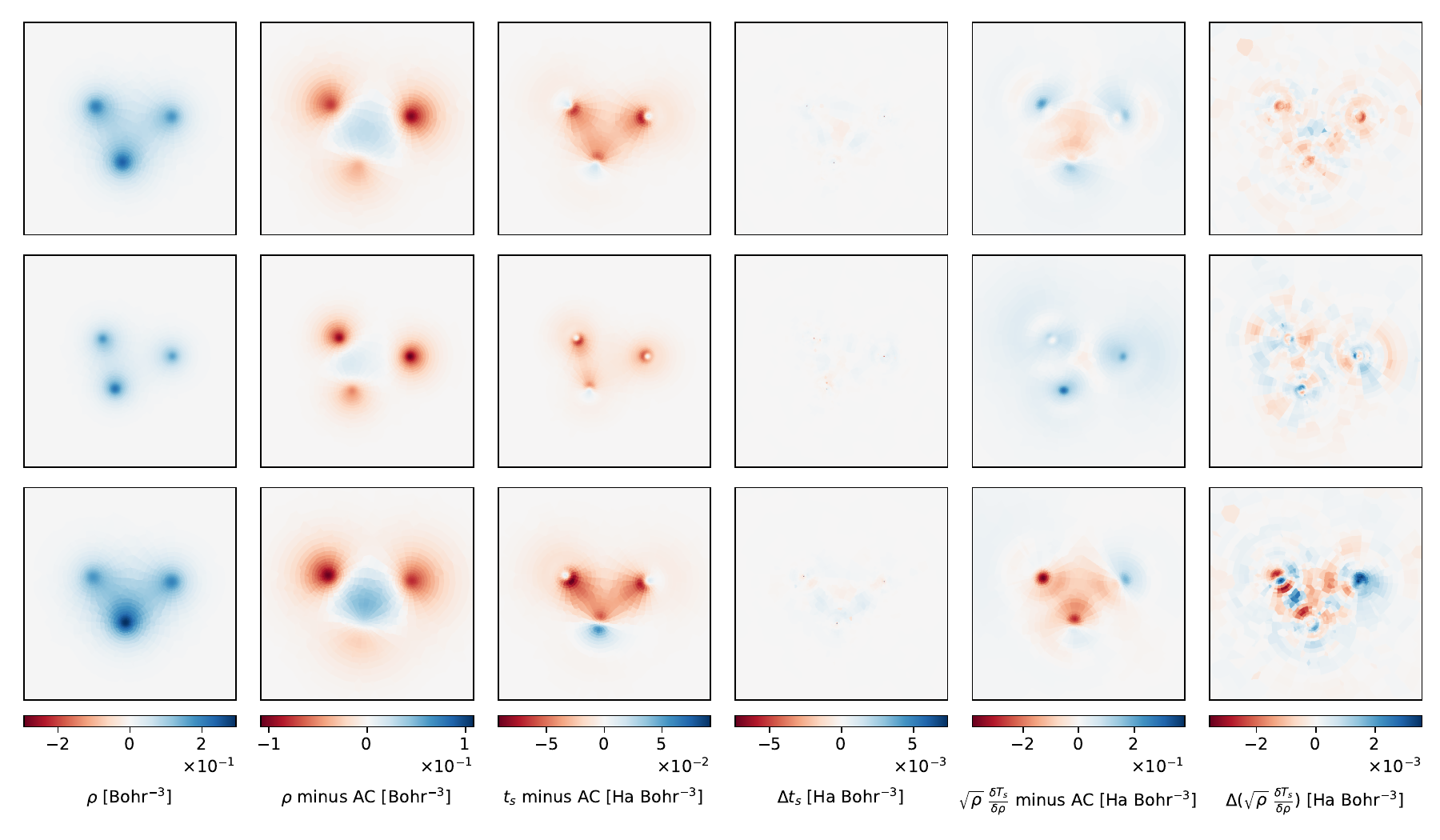}
\caption{\label{fig:grid} Slices through input, prediction, and error for \ce{H3+}, on three test samples. From left to right: Electron density, electron density minus AC, predicted kinetic energy density minus AC, error of the electron density, predicted kinetic potential times square root of electron density and lastly its error.}
\end{figure*}
\begin{table*}[]
\caption{Energy mean absolute error for our models (KineticNet) and classical functionals on test sets consisting of KS solution densities (for varying $v_{\text{ext}}$).\label{tab:validation_energy}}
\sisetup{table-format = 1.1e2}
\begin{tabular}{r|S|S|S|S|S|S}
$|\Delta T_s| [mHa]$ & \ce{He} & \ce{H2} & \ce{H3+} & \ce{HF} & \ce{Ne2} & \ce{H2O} \\ \hline
KineticNet        & 1.3e-1   & 6.4e-1   & 4.7e-1   &  2.0e0   &  1.0e1   &  1.3e0    \\ 
TF                & 2.9e2     & 2.1e2     & 2.8e2      & 9.1e3    & 2.2e4    & 6.9e3     \\ 
vW                &0       &0       &0        & 2.6e4   & 7.8e4    & 1.9e4     \\ 
MGE2              & 7.9e1      & 1.0e2     & 1.6e2      & 7.8e2     & 2.6e3     & 8.1e2  
\end{tabular}
\end{table*}
For all data sets the mean absolute error (MAE) of the predicted kinetic energy over 100 samples from the test set falls below the threshold of chemical accuracy,  1 mHa per electron, see table \ref{tab:validation_energy}. 
We compare these results to three classical approximations of the kinetic energy functional, the first-order Thomas-Fermi (TF) functional, the second order von Weizäcker (vW) correction and the MGE2 functional which is a linear combination of the two former approximations and which performed best in the extensive comparison by \citeauthor{fujinami2020orbital}\cite{fujinami2020orbital}. 
The superiority of the ML functional in this metric is very obvious as it outperforms the classical approximations by more than two orders of magnitude throughout, however with one exception:
The vW functional is exact for two electron systems, hence its MAE is zero for \ce{He}, \ce{H2} and \ce{H3+}. One could argue that the learning task for the ML model in these cases is also much easier for the ML functional, as the semilocal expression of the vW functional is already exact, but on one hand our model by construction cannot simply reproduce this term, and on the other hand we demonstrate a similar accuracy (per electron) on the bigger systems, where vW alone fails spectacularly. 

For \ce{HF} and \ce {Ne2} we additionally demonstrate that our model is accurate enough to model chemical bonds (or the absence thereof) by evaluating it on the KS ground states for varying inter-atomic distances, and plotting the resulting dissociation curves in figures \ref{fig:hf_dissociation} and \ref{fig:ne_dissociation}. Note that none of the geometries, nor any ground states (without a perturbed external potential) were part of the training sets. 
For \ce{H3+}, a comparison between predicted and target kinetic energies is shown in figure \ref{fig:scatter_ekin_h3+}.

\subsection{Density optimization}
\begin{figure}
\includegraphics[width=8cm]{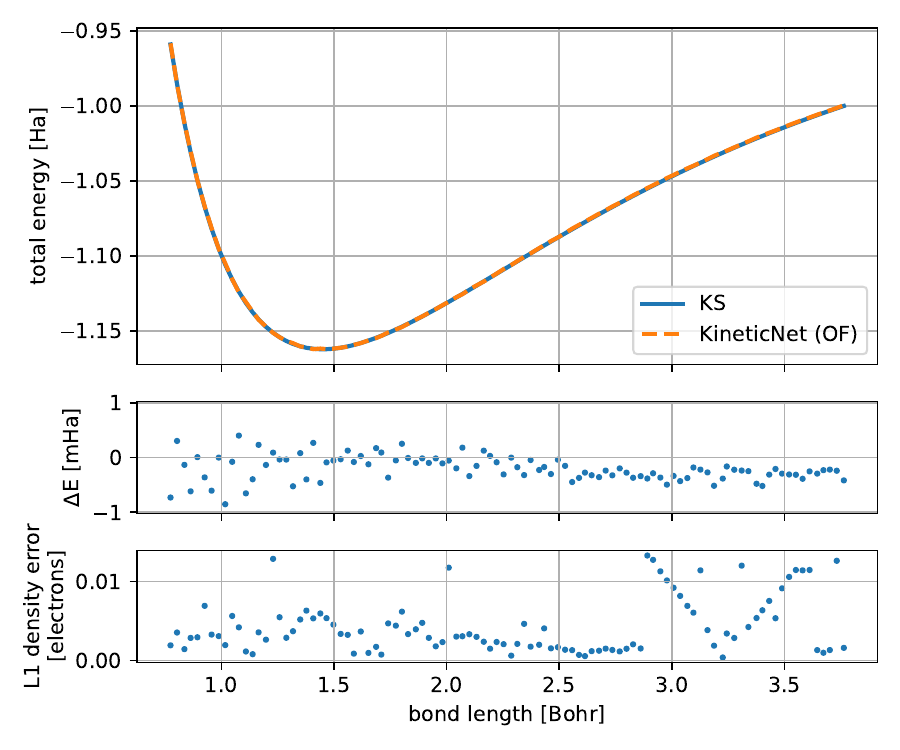}%
\caption{\label{fig:H2_dissociation} Total ground state energy of $H_2$ at different bond lengths, as computed by KS-DFT as well as OF-DFT using our machine learned functionals.}
\end{figure}
\begin{table*}[t]
\caption{Density optimization results for two electron systems using our ML functionals as well as classical functionals. Note that the vW functional is exact for these systems, the small deviations are due to the limited steps and finite convergence threshold in our OF-DFT implementation.\label{tab:dens_optim}}

\begin{tabular}{l|c|S[table-format=3.2]S|S[table-format=3.2]S|S[table-format=3.2]S}
     & \multicolumn{1}{l|}{} & \multicolumn{2}{c|}{$\mathbf{v}_{\text{ext}}=0$}          & \multicolumn{2}{c|}{$\mathbf{v}_{\text{ext}}$ from test} & \multicolumn{2}{c}{with solvation model}         \\
     & Data set               & {$\Delta E$ [mHa]} & {$\|\rho-\rho_{\text{KS}}\|_1$} & {$\Delta E$ [mHa]}    & {$\|\rho-\rho_{\text{KS}}\|_1$}   & {$\Delta E$ [mHa]} & {$\|\rho-\rho_{\text{KS}}\|_1$} \\ \hline
             & \ce{He}    & \cc0.04          & \cc0.0002                     & \cc0.12             & \cc0.0014                       & \cc0.04          & \cc0.0002       \\
KineticNet   & \ce{H2}    & \cc0.32          & \cc0.0039                     & \cc0.37             & \cc0.0106                       & \cc0.32          & \cc0.0040       \\
             & \ce{H3+}   & \cc0.42          & \cc0.0056                     & \cc0.44             & \cc0.0104                       & \cc0.43          & \cc0.0054   \\ \hline
     & \ce{He}   & \cc305.30  & \cc0.1282  & \cc306.50  & \cc0.1306  & \cc305.30  & \cc0.1282   \\
TF   & \ce{H2}   & \cc394.37  & \cc1.5073  & \cc750.29  & \cc1.5984  & \cc394.59  & \cc1.5098   \\
     & \ce{H3+}  & \cc685.16  & \cc1.2837  & \cc735.07  & \cc1.3759  & \cc691.76  & \cc1.3063   \\ \hline
     & \ce{He}   & \cc0.00    & \cc0.0000  & \cc0.00    & \cc0.0003  & \cc0.00    & \cc0.0000   \\
vW   & \ce{H2}   & \cc0.01    & \cc0.0001  & \cc0.01    & \cc0.0005  & \cc0.01    & \cc0.0001   \\
     & \ce{H3+}  & \cc0.02    & \cc0.0007  & \cc0.02    & \cc0.0008  & \cc0.02    & \cc0.0007   \\ \hline
     & \ce{He}   & \cc76.73   & \cc0.1367  & \cc77.63   & \cc0.1391  & \cc76.73   & \cc0.1367   \\
MGE2 & \ce{H2}   & \cc129.53  & \cc0.5992  & \cc332.39  & \cc1.2301  & \cc129.15  & \cc0.6017   \\
     & \ce{H3+}  & \cc371.04  & \cc1.0251  & \cc415.85  & \cc1.1342  & \cc373.60  & \cc1.0460   \\ \hline
     & \ce{He}   & \cc5.45    & \cc0.0878  & \cc6.24    & \cc0.0900  & \cc5.45    & \cc0.0878   \\
LC   & \ce{H2}   & \cc166.72  & \cc0.7546  & \cc516.21  & \cc1.4945  & \cc166.31  & \cc0.7573   \\
     & \ce{H3+}  & \cc485.04  & \cc1.1881  & \cc541.69  & \cc1.3018  & \cc490.12  & \cc1.2152  
\end{tabular}
\end{table*}
The results of applying our machine learned functionals in OF-DFT are summarized in table \ref{tab:dens_optim}.
We evaluated each of our models on \correction{200} geometries from the corresponding test set (except of course for \ce{He}, where only a single geometry is available), setting the SLSQP convergence threshold to \correction{$10^{-6}$ hartree} and allowing a maximum of 100 steps. 
While we always observe convergence using the classical functional approximations, in the few cases when no convergence is reached using our model, we evaluate the best solution obtained so far.
To quantify the results, we compute the mean over the absolute energy errors, as well as the L1 density deviation
\begin{align}\label{eq:L1-density-deviation}
    \|\rho-\rho_{\text{KS}}\|_1 = \int |\rho(\mathbf{r}) - \rho_{\text{KS}}(\mathbf{r})| \, \text{d}\mathbf{r}
\end{align}
between the KS density $\rho_{\text{KS}}$ and the result of the OF calculation $\rho$.

For \ce{He}, \ce{H2} and \ce{H3+}, we obtain errors of less than 1 mHa and L1 density deviations on the order of $10^{-2}$ electrons, see table \ref{tab:dens_optim}.
This is more than precise enough to correctly model the \ce{H2} bond, see figure \ref{fig:H2_dissociation}.

Furthermore, the way our learned functionals generalize allows us to apply them in different settings: Just as KS-DFT during data generation, we can apply orbital free density optimization on molecules in the presence of an additional external potential. 
For this, we use potentials from the test set and observe that the accuracy of our model is still good, see middle two columns in table \ref{tab:dens_optim}.

We can also apply solvation models that simulate a chemical environment by a density-dependent contribution to the external potential. To this end we employ the ddCOSMO solvation model \cite{cances2013domain,lipparini2013fast,lipparini2014quantum} as implemented in pyscf with default parameters, i.e.~simulating a solution in water, see the two rightmost columns of table \ref{tab:dens_optim}.

Note that none of these modifications would have been possible if we took a very direct black-box ML approach of e.g.~directly predicting the ground-state electron density.

The reason why we only present density optimization results for the two electron systems \ce{He}, \ce{H2} and \ce{H3+} is that only for those there is an exact correspondence between the possible KS densities and the Ansatz we use in our OF calculations (eq.\ref{eq:of-dft-ansatz}, for details see appendix C). 
Hence, for these systems densities close to the KS ground state are obtainable. 
On the other hand, in the cc-pVDZ basis that we are using, it is impossible to model densities close to the KS ground state using the OF Ansatz, even fitting the coefficients to best mimic the KS density leads to an L1 deviation of multiple electrons.
So while OF-DFT calculations using our learned functionals sometimes converge for these larger systems, either a sufficiently larger basis, maybe optimized for this application, or an entirely different Ansatz are required to reach quantitatively interesting results.

\section{Conclusion}\label{sec:conclusion}

We present KineticNet, a new equivariant machine learning model adapted for the prediction of molecular properties on quadrature grids. 
Using the electron density on the grid and the positions of all nuclei as input, it can successfully predict the corresponding non-interacting kinetic energy density for a variety of systems such as \ce{HF}, \ce{H2O} and \ce{Ne2}. 
The new functional correctly describes both chemical bonding (as well as the absence of it in \ce{Ne2}).
We offer proof of principle that this architecture can predict the kinetic potential with sufficient accuracy to allow actual OF-DFT density optimization to reach the respective KS-DFT ground state for the model systems \ce{H2}, \ce{H3+} and \ce{Ne}.
Additionally, we show that the generation of varied training data, by invoking fundamental concepts of DFT, allows training models with \correction{little} overfitting which generalize over densities arising in the presence of different external potentials.
This also includes simple solvent models such as ddCOSMO which can be applied out of the box without any additional retraining.
\correction{

When designing KineticNet, we were not attempting to find the simplest possible architecture to produce meaningful results on the small molecules used here, but one which is conceptually capable of generalizing over different molecular compositions as well as system sizes. Extensive hyperparameter optimization
can be expected to pay off, and is part of future work. }

Given \correction{the} encouraging results, the next step is to generalize the entire workflow to afford density optimization for more than two electrons. We conjecture that the principal obstacle in the current setup is that the KS-DFT ground state cannot be represented by our combination of basis and description of the density.
In response, we are now working on a specialized OF-DFT basis set which hopefully overcomes this limitation and allows for chemically accurate OF-DFT calculations on systems of unprecedented size.
\correction{

In this work we opted for a non-variational approach, estimating the potential with an independent network and not as a derivative of the energy, in an attempt to make the most of the available GPU memory and training time. For future work, it looks promising to revisit the variational approach, as chemically correct model gradients will become important for geometry optimization and molecular dynamics.}

\subsection*{Acknowledgements}
We would like to thank Christof Gehrig for his help in achieving the first successful density optimization, in particular by suggesting the use of the product of the kinetic potential and the square root of the electron density as a target. \correction{The constructive criticism of two anonymous reviewers helped a lot in clarifying the present approach and its limitations.} 
This work is supported by Deutsche Forschungsgemeinschaft (DFG) under Germany’s Excellence Strategy EXC-2181/1 - 390900948 (the Heidelberg STRUCTURES Excellence Cluster), as well as by Klaus Tschira Stiftung gGmbH in the framework of the SIMPLAIX consortium.
T.K.~and A.D.\ acknowledge support by the state of Baden-Württemberg through bwHPC
and the German Research Foundation (DFG) through grant no INST 40/575-1 FUGG (JUSTUS 2 cluster). We also thank the SFB 1249 for funding this research.

\section*{Author declarations}
\subsection*{Conflict of Interest}
The authors have no conflicts to disclose.

\subsection*{Author Contributions}
\noindent\mbox{\textbf{Roman Remme}}: Data Curation (equal); Methodology (lead); Software (equal); Writing/Original Draft Preparation (lead); Writing/Review \& Editing (equal).
\mbox{\textbf{Tobias Kaczun}}: Data Curation (equal); Methodology (supporting); Software (equal); Writing/Original Draft Preparation (supporting); Writing/Review \& Editing (equal).
\mbox{\textbf{Maximilian Scheurer}}: Software (equal); Writing/Review \& Editing (supporting).
\mbox{\textbf{Andreas Dreuw}}: Conceptualization (supporting); Supervision (equal); Writing/Review \& Editing (supporting).
\mbox{\textbf{Fred A.~Hamprecht}}: Conceptualization (lead); Supervision (equal); Writing/Review \& Editing (supporting).

\section{Data availability statement}
The data that support the findings of this study are available from the corresponding author upon reasonable request.

\section*{References}
\bibliography{aipmain}%

\begin{thebibliography}{37}%
\makeatletter
\providecommand \@ifxundefined [1]{%
 \@ifx{#1\undefined}
}%
\providecommand \@ifnum [1]{%
 \ifnum #1\expandafter \@firstoftwo
 \else \expandafter \@secondoftwo
 \fi
}%
\providecommand \@ifx [1]{%
 \ifx #1\expandafter \@firstoftwo
 \else \expandafter \@secondoftwo
 \fi
}%
\providecommand \natexlab [1]{#1}%
\providecommand \enquote  [1]{``#1''}%
\providecommand \bibnamefont  [1]{#1}%
\providecommand \bibfnamefont [1]{#1}%
\providecommand \citenamefont [1]{#1}%
\providecommand \href@noop [0]{\@secondoftwo}%
\providecommand \href [0]{\begingroup \@sanitize@url \@href}%
\providecommand \@href[1]{\@@startlink{#1}\@@href}%
\providecommand \@@href[1]{\endgroup#1\@@endlink}%
\providecommand \@sanitize@url [0]{\catcode `\\12\catcode `\$12\catcode
  `\&12\catcode `\#12\catcode `\^12\catcode `\_12\catcode `\%12\relax}%
\providecommand \@@startlink[1]{}%
\providecommand \@@endlink[0]{}%
\providecommand \url  [0]{\begingroup\@sanitize@url \@url }%
\providecommand \@url [1]{\endgroup\@href {#1}{\urlprefix }}%
\providecommand \urlprefix  [0]{URL }%
\providecommand \Eprint [0]{\href }%
\providecommand \doibase [0]{http://dx.doi.org/}%
\providecommand \selectlanguage [0]{\@gobble}%
\providecommand \bibinfo  [0]{\@secondoftwo}%
\providecommand \bibfield  [0]{\@secondoftwo}%
\providecommand \translation [1]{[#1]}%
\providecommand \BibitemOpen [0]{}%
\providecommand \bibitemStop [0]{}%
\providecommand \bibitemNoStop [0]{.\EOS\space}%
\providecommand \EOS [0]{\spacefactor3000\relax}%
\providecommand \BibitemShut  [1]{\csname bibitem#1\endcsname}%
\let\auto@bib@innerbib\@empty
\bibitem [{\citenamefont {Parr}\ and\ \citenamefont
  {Yang}(1995)}]{parr1995density}%
  \BibitemOpen
  \bibfield  {author} {\bibinfo {author} {\bibfnamefont {R.~G.}\ \bibnamefont
  {Parr}}\ and\ \bibinfo {author} {\bibfnamefont {W.}~\bibnamefont {Yang}},\
  }\bibfield  {title} {\enquote {\bibinfo {title} {Density-functional theory of
  the electronic structure of molecules},}\ }\href@noop {} {\bibfield
  {journal} {\bibinfo  {journal} {Annual review of physical chemistry}\
  }\textbf {\bibinfo {volume} {46}},\ \bibinfo {pages} {701--728} (\bibinfo
  {year} {1995})}\BibitemShut {NoStop}%
\bibitem [{\citenamefont {Kirkpatrick}\ \emph {et~al.}(2021)\citenamefont
  {Kirkpatrick}, \citenamefont {McMorrow}, \citenamefont {Turban},
  \citenamefont {Gaunt}, \citenamefont {Spencer}, \citenamefont {Matthews},
  \citenamefont {Obika}, \citenamefont {Thiry}, \citenamefont {Fortunato},
  \citenamefont {Pfau}, \citenamefont {Castellanos}, \citenamefont {Petersen},
  \citenamefont {Nelson}, \citenamefont {Kohli}, \citenamefont {Mori-Sánchez},
  \citenamefont {Hassabis},\ and\ \citenamefont
  {Cohen}}]{kirkpatrick2021pushing}%
  \BibitemOpen
  \bibfield  {author} {\bibinfo {author} {\bibfnamefont {J.}~\bibnamefont
  {Kirkpatrick}}, \bibinfo {author} {\bibfnamefont {B.}~\bibnamefont
  {McMorrow}}, \bibinfo {author} {\bibfnamefont {D.~H.~P.}\ \bibnamefont
  {Turban}}, \bibinfo {author} {\bibfnamefont {A.~L.}\ \bibnamefont {Gaunt}},
  \bibinfo {author} {\bibfnamefont {J.~S.}\ \bibnamefont {Spencer}}, \bibinfo
  {author} {\bibfnamefont {A.~G. D.~G.}\ \bibnamefont {Matthews}}, \bibinfo
  {author} {\bibfnamefont {A.}~\bibnamefont {Obika}}, \bibinfo {author}
  {\bibfnamefont {L.}~\bibnamefont {Thiry}}, \bibinfo {author} {\bibfnamefont
  {M.}~\bibnamefont {Fortunato}}, \bibinfo {author} {\bibfnamefont
  {D.}~\bibnamefont {Pfau}}, \bibinfo {author} {\bibfnamefont {L.~R.}\
  \bibnamefont {Castellanos}}, \bibinfo {author} {\bibfnamefont
  {S.}~\bibnamefont {Petersen}}, \bibinfo {author} {\bibfnamefont {A.~W.~R.}\
  \bibnamefont {Nelson}}, \bibinfo {author} {\bibfnamefont {P.}~\bibnamefont
  {Kohli}}, \bibinfo {author} {\bibfnamefont {P.}~\bibnamefont
  {Mori-Sánchez}}, \bibinfo {author} {\bibfnamefont {D.}~\bibnamefont
  {Hassabis}}, \ and\ \bibinfo {author} {\bibfnamefont {A.~J.}\ \bibnamefont
  {Cohen}},\ }\bibfield  {title} {\enquote {\bibinfo {title} {Pushing the
  frontiers of density functionals by solving the fractional electron
  problem},}\ }\href {\doibase 10.1126/science.abj6511} {\bibfield  {journal}
  {\bibinfo  {journal} {Science}\ }\textbf {\bibinfo {volume} {374}},\ \bibinfo
  {pages} {1385–1389} (\bibinfo {year} {2021})}\BibitemShut {NoStop}%
\bibitem [{\citenamefont {Nagai}, \citenamefont {Akashi},\ and\ \citenamefont
  {Sugino}(2020)}]{nagai2020completing}%
  \BibitemOpen
  \bibfield  {author} {\bibinfo {author} {\bibfnamefont {R.}~\bibnamefont
  {Nagai}}, \bibinfo {author} {\bibfnamefont {R.}~\bibnamefont {Akashi}}, \
  and\ \bibinfo {author} {\bibfnamefont {O.}~\bibnamefont {Sugino}},\
  }\bibfield  {title} {\enquote {\bibinfo {title} {Completing density
  functional theory by machine learning hidden messages from molecules},}\
  }\href@noop {} {\bibfield  {journal} {\bibinfo  {journal} {npj Computational
  Materials}\ }\textbf {\bibinfo {volume} {6}},\ \bibinfo {pages} {1--8}
  (\bibinfo {year} {2020})}\BibitemShut {NoStop}%
\bibitem [{\citenamefont {Bystrom}\ and\ \citenamefont
  {Kozinsky}(2022)}]{bystrom2022cider}%
  \BibitemOpen
  \bibfield  {author} {\bibinfo {author} {\bibfnamefont {K.}~\bibnamefont
  {Bystrom}}\ and\ \bibinfo {author} {\bibfnamefont {B.}~\bibnamefont
  {Kozinsky}},\ }\bibfield  {title} {\enquote {\bibinfo {title} {Cider: An
  expressive, nonlocal feature set for machine learning density functionals
  with exact constraints},}\ }\href@noop {} {\bibfield  {journal} {\bibinfo
  {journal} {Journal of Chemical Theory and Computation}\ }\textbf {\bibinfo
  {volume} {18}},\ \bibinfo {pages} {2180--2192} (\bibinfo {year}
  {2022})}\BibitemShut {NoStop}%
\bibitem [{\citenamefont {Dick}\ and\ \citenamefont
  {Fernandez-Serra}(2020)}]{dick2020machine}%
  \BibitemOpen
  \bibfield  {author} {\bibinfo {author} {\bibfnamefont {S.}~\bibnamefont
  {Dick}}\ and\ \bibinfo {author} {\bibfnamefont {M.}~\bibnamefont
  {Fernandez-Serra}},\ }\bibfield  {title} {\enquote {\bibinfo {title} {Machine
  learning accurate exchange and correlation functionals of the electronic
  density},}\ }\href@noop {} {\bibfield  {journal} {\bibinfo  {journal} {Nature
  communications}\ }\textbf {\bibinfo {volume} {11}},\ \bibinfo {pages} {1--10}
  (\bibinfo {year} {2020})}\BibitemShut {NoStop}%
\bibitem [{\citenamefont {Mardirossian}\ and\ \citenamefont
  {Head-Gordon}(2014)}]{mardirossian2014omegab97x}%
  \BibitemOpen
  \bibfield  {author} {\bibinfo {author} {\bibfnamefont {N.}~\bibnamefont
  {Mardirossian}}\ and\ \bibinfo {author} {\bibfnamefont {M.}~\bibnamefont
  {Head-Gordon}},\ }\bibfield  {title} {\enquote {\bibinfo {title}
  {$\omega$b97x-v: A 10-parameter, range-separated hybrid, generalized gradient
  approximation density functional with nonlocal correlation, designed by a
  survival-of-the-fittest strategy},}\ }\href@noop {} {\bibfield  {journal}
  {\bibinfo  {journal} {Physical Chemistry Chemical Physics}\ }\textbf
  {\bibinfo {volume} {16}},\ \bibinfo {pages} {9904--9924} (\bibinfo {year}
  {2014})}\BibitemShut {NoStop}%
\bibitem [{\citenamefont {Schmidt}, \citenamefont {Benavides-Riveros},\ and\
  \citenamefont {Marques}(2019)}]{schmidt2019machine}%
  \BibitemOpen
  \bibfield  {author} {\bibinfo {author} {\bibfnamefont {J.}~\bibnamefont
  {Schmidt}}, \bibinfo {author} {\bibfnamefont {C.~L.}\ \bibnamefont
  {Benavides-Riveros}}, \ and\ \bibinfo {author} {\bibfnamefont {M.~A.}\
  \bibnamefont {Marques}},\ }\bibfield  {title} {\enquote {\bibinfo {title}
  {Machine learning the physical nonlocal exchange--correlation functional of
  density-functional theory},}\ }\href@noop {} {\bibfield  {journal} {\bibinfo
  {journal} {The journal of physical chemistry letters}\ }\textbf {\bibinfo
  {volume} {10}},\ \bibinfo {pages} {6425--6431} (\bibinfo {year}
  {2019})}\BibitemShut {NoStop}%
\bibitem [{\citenamefont {Snyder}\ \emph {et~al.}(2013)\citenamefont {Snyder},
  \citenamefont {Rupp}, \citenamefont {Hansen}, \citenamefont {Blooston},
  \citenamefont {M{\"u}ller},\ and\ \citenamefont {Burke}}]{snyder2013orbital}%
  \BibitemOpen
  \bibfield  {author} {\bibinfo {author} {\bibfnamefont {J.~C.}\ \bibnamefont
  {Snyder}}, \bibinfo {author} {\bibfnamefont {M.}~\bibnamefont {Rupp}},
  \bibinfo {author} {\bibfnamefont {K.}~\bibnamefont {Hansen}}, \bibinfo
  {author} {\bibfnamefont {L.}~\bibnamefont {Blooston}}, \bibinfo {author}
  {\bibfnamefont {K.-R.}\ \bibnamefont {M{\"u}ller}}, \ and\ \bibinfo {author}
  {\bibfnamefont {K.}~\bibnamefont {Burke}},\ }\bibfield  {title} {\enquote
  {\bibinfo {title} {Orbital-free bond breaking via machine learning},}\
  }\href@noop {} {\bibfield  {journal} {\bibinfo  {journal} {The Journal of
  chemical physics}\ }\textbf {\bibinfo {volume} {139}},\ \bibinfo {pages}
  {224104} (\bibinfo {year} {2013})}\BibitemShut {NoStop}%
\bibitem [{\citenamefont {Meyer}, \citenamefont {Weichselbaum},\ and\
  \citenamefont {Hauser}(2020)}]{meyer2020machine}%
  \BibitemOpen
  \bibfield  {author} {\bibinfo {author} {\bibfnamefont {R.}~\bibnamefont
  {Meyer}}, \bibinfo {author} {\bibfnamefont {M.}~\bibnamefont {Weichselbaum}},
  \ and\ \bibinfo {author} {\bibfnamefont {A.~W.}\ \bibnamefont {Hauser}},\
  }\bibfield  {title} {\enquote {\bibinfo {title} {Machine learning approaches
  toward orbital-free density functional theory: Simultaneous training on the
  kinetic energy density functional and its functional derivative},}\
  }\href@noop {} {\bibfield  {journal} {\bibinfo  {journal} {Journal of
  chemical theory and computation}\ }\textbf {\bibinfo {volume} {16}},\
  \bibinfo {pages} {5685--5694} (\bibinfo {year} {2020})}\BibitemShut {NoStop}%
\bibitem [{\citenamefont {Saidaoui}\ \emph {et~al.}(2020)\citenamefont
  {Saidaoui}, \citenamefont {Kais}, \citenamefont {Rashkeev},\ and\
  \citenamefont {Alharbi}}]{saidaoui2020direct}%
  \BibitemOpen
  \bibfield  {author} {\bibinfo {author} {\bibfnamefont {H.}~\bibnamefont
  {Saidaoui}}, \bibinfo {author} {\bibfnamefont {S.}~\bibnamefont {Kais}},
  \bibinfo {author} {\bibfnamefont {S.}~\bibnamefont {Rashkeev}}, \ and\
  \bibinfo {author} {\bibfnamefont {F.}~\bibnamefont {Alharbi}},\ }\bibfield
  {title} {\enquote {\bibinfo {title} {Direct scheme calculation of the kinetic
  energy functional derivative using machine learning},}\ }\href@noop {}
  {\bibfield  {journal} {\bibinfo  {journal} {arXiv preprint arXiv:2003.00876}\
  } (\bibinfo {year} {2020})}\BibitemShut {NoStop}%
\bibitem [{\citenamefont {Ghasemi}\ and\ \citenamefont
  {K{\"u}hne}(2021)}]{ghasemi2021artificial}%
  \BibitemOpen
  \bibfield  {author} {\bibinfo {author} {\bibfnamefont {S.~A.}\ \bibnamefont
  {Ghasemi}}\ and\ \bibinfo {author} {\bibfnamefont {T.~D.}\ \bibnamefont
  {K{\"u}hne}},\ }\bibfield  {title} {\enquote {\bibinfo {title} {Artificial
  neural networks for the kinetic energy functional of non-interacting
  fermions},}\ }\href@noop {} {\bibfield  {journal} {\bibinfo  {journal} {The
  Journal of Chemical Physics}\ }\textbf {\bibinfo {volume} {154}},\ \bibinfo
  {pages} {074107} (\bibinfo {year} {2021})}\BibitemShut {NoStop}%
\bibitem [{\citenamefont {Golub}\ and\ \citenamefont
  {Manzhos}(2019)}]{golub2019kinetic}%
  \BibitemOpen
  \bibfield  {author} {\bibinfo {author} {\bibfnamefont {P.}~\bibnamefont
  {Golub}}\ and\ \bibinfo {author} {\bibfnamefont {S.}~\bibnamefont
  {Manzhos}},\ }\bibfield  {title} {\enquote {\bibinfo {title} {Kinetic energy
  densities based on the fourth order gradient expansion: performance in
  different classes of materials and improvement via machine learning},}\
  }\href@noop {} {\bibfield  {journal} {\bibinfo  {journal} {Physical Chemistry
  Chemical Physics}\ }\textbf {\bibinfo {volume} {21}},\ \bibinfo {pages}
  {378--395} (\bibinfo {year} {2019})}\BibitemShut {NoStop}%
\bibitem [{\citenamefont {Seino}\ \emph {et~al.}(2018)\citenamefont {Seino},
  \citenamefont {Kageyama}, \citenamefont {Fujinami}, \citenamefont {Ikabata},\
  and\ \citenamefont {Nakai}}]{seino2018semi}%
  \BibitemOpen
  \bibfield  {author} {\bibinfo {author} {\bibfnamefont {J.}~\bibnamefont
  {Seino}}, \bibinfo {author} {\bibfnamefont {R.}~\bibnamefont {Kageyama}},
  \bibinfo {author} {\bibfnamefont {M.}~\bibnamefont {Fujinami}}, \bibinfo
  {author} {\bibfnamefont {Y.}~\bibnamefont {Ikabata}}, \ and\ \bibinfo
  {author} {\bibfnamefont {H.}~\bibnamefont {Nakai}},\ }\bibfield  {title}
  {\enquote {\bibinfo {title} {Semi-local machine-learned kinetic energy
  density functional with third-order gradients of electron density},}\
  }\href@noop {} {\bibfield  {journal} {\bibinfo  {journal} {The Journal of
  chemical physics}\ }\textbf {\bibinfo {volume} {148}},\ \bibinfo {pages}
  {241705} (\bibinfo {year} {2018})}\BibitemShut {NoStop}%
\bibitem [{\citenamefont {Fujinami}\ \emph {et~al.}(2020)\citenamefont
  {Fujinami}, \citenamefont {Kageyama}, \citenamefont {Seino}, \citenamefont
  {Ikabata},\ and\ \citenamefont {Nakai}}]{fujinami2020orbital}%
  \BibitemOpen
  \bibfield  {author} {\bibinfo {author} {\bibfnamefont {M.}~\bibnamefont
  {Fujinami}}, \bibinfo {author} {\bibfnamefont {R.}~\bibnamefont {Kageyama}},
  \bibinfo {author} {\bibfnamefont {J.}~\bibnamefont {Seino}}, \bibinfo
  {author} {\bibfnamefont {Y.}~\bibnamefont {Ikabata}}, \ and\ \bibinfo
  {author} {\bibfnamefont {H.}~\bibnamefont {Nakai}},\ }\bibfield  {title}
  {\enquote {\bibinfo {title} {Orbital-free density functional theory
  calculation applying semi-local machine-learned kinetic energy density
  functional and kinetic potential},}\ }\href@noop {} {\bibfield  {journal}
  {\bibinfo  {journal} {Chemical Physics Letters}\ }\textbf {\bibinfo {volume}
  {748}},\ \bibinfo {pages} {137358} (\bibinfo {year} {2020})}\BibitemShut
  {NoStop}%
\bibitem [{\citenamefont {Ryczko}\ \emph {et~al.}(2022)\citenamefont {Ryczko},
  \citenamefont {Wetzel}, \citenamefont {Melko},\ and\ \citenamefont
  {Tamblyn}}]{ryczko2022toward}%
  \BibitemOpen
  \bibfield  {author} {\bibinfo {author} {\bibfnamefont {K.}~\bibnamefont
  {Ryczko}}, \bibinfo {author} {\bibfnamefont {S.~J.}\ \bibnamefont {Wetzel}},
  \bibinfo {author} {\bibfnamefont {R.~G.}\ \bibnamefont {Melko}}, \ and\
  \bibinfo {author} {\bibfnamefont {I.}~\bibnamefont {Tamblyn}},\ }\bibfield
  {title} {\enquote {\bibinfo {title} {Toward orbital-free density functional
  theory with small data sets and deep learning},}\ }\href@noop {} {\bibfield
  {journal} {\bibinfo  {journal} {Journal of Chemical Theory and Computation}\
  }\textbf {\bibinfo {volume} {18}},\ \bibinfo {pages} {1122--1128} (\bibinfo
  {year} {2022})}\BibitemShut {NoStop}%
\bibitem [{\citenamefont {Imoto}, \citenamefont {Imada},\ and\ \citenamefont
  {Oshiyama}(2021)}]{imoto2021order}%
  \BibitemOpen
  \bibfield  {author} {\bibinfo {author} {\bibfnamefont {F.}~\bibnamefont
  {Imoto}}, \bibinfo {author} {\bibfnamefont {M.}~\bibnamefont {Imada}}, \ and\
  \bibinfo {author} {\bibfnamefont {A.}~\bibnamefont {Oshiyama}},\ }\bibfield
  {title} {\enquote {\bibinfo {title} {Order-n orbital-free density-functional
  calculations with machine learning of functional derivatives for
  semiconductors and metals},}\ }\href@noop {} {\bibfield  {journal} {\bibinfo
  {journal} {Physical Review Research}\ }\textbf {\bibinfo {volume} {3}},\
  \bibinfo {pages} {033198} (\bibinfo {year} {2021})}\BibitemShut {NoStop}%
\bibitem [{\citenamefont {Thomas}\ \emph {et~al.}(2018)\citenamefont {Thomas},
  \citenamefont {Smidt}, \citenamefont {Kearnes}, \citenamefont {Yang},
  \citenamefont {Li}, \citenamefont {Kohlhoff},\ and\ \citenamefont
  {Riley}}]{thomas2018tensor}%
  \BibitemOpen
  \bibfield  {author} {\bibinfo {author} {\bibfnamefont {N.}~\bibnamefont
  {Thomas}}, \bibinfo {author} {\bibfnamefont {T.}~\bibnamefont {Smidt}},
  \bibinfo {author} {\bibfnamefont {S.}~\bibnamefont {Kearnes}}, \bibinfo
  {author} {\bibfnamefont {L.}~\bibnamefont {Yang}}, \bibinfo {author}
  {\bibfnamefont {L.}~\bibnamefont {Li}}, \bibinfo {author} {\bibfnamefont
  {K.}~\bibnamefont {Kohlhoff}}, \ and\ \bibinfo {author} {\bibfnamefont
  {P.}~\bibnamefont {Riley}},\ }\bibfield  {title} {\enquote {\bibinfo {title}
  {Tensor field networks: Rotation-and translation-equivariant neural networks
  for 3d point clouds},}\ }\href@noop {} {\bibfield  {journal} {\bibinfo
  {journal} {arXiv preprint arXiv:1802.08219}\ } (\bibinfo {year}
  {2018})}\BibitemShut {NoStop}%
\bibitem [{\citenamefont {Geiger}\ and\ \citenamefont
  {Smidt}(2022)}]{geiger2022e3nn}%
  \BibitemOpen
  \bibfield  {author} {\bibinfo {author} {\bibfnamefont {M.}~\bibnamefont
  {Geiger}}\ and\ \bibinfo {author} {\bibfnamefont {T.}~\bibnamefont {Smidt}},\
  }\bibfield  {title} {\enquote {\bibinfo {title} {e3nn: Euclidean neural
  networks},}\ }\href@noop {} {\bibfield  {journal} {\bibinfo  {journal} {arXiv
  preprint arXiv:2207.09453}\ } (\bibinfo {year} {2022})}\BibitemShut {NoStop}%
\bibitem [{\citenamefont {Grisafi}\ \emph {et~al.}(2018)\citenamefont
  {Grisafi}, \citenamefont {Wilkins}, \citenamefont {Cs{\'a}nyi},\ and\
  \citenamefont {Ceriotti}}]{grisafi2018symmetry}%
  \BibitemOpen
  \bibfield  {author} {\bibinfo {author} {\bibfnamefont {A.}~\bibnamefont
  {Grisafi}}, \bibinfo {author} {\bibfnamefont {D.~M.}\ \bibnamefont
  {Wilkins}}, \bibinfo {author} {\bibfnamefont {G.}~\bibnamefont {Cs{\'a}nyi}},
  \ and\ \bibinfo {author} {\bibfnamefont {M.}~\bibnamefont {Ceriotti}},\
  }\bibfield  {title} {\enquote {\bibinfo {title} {Symmetry-adapted machine
  learning for tensorial properties of atomistic systems},}\ }\href@noop {}
  {\bibfield  {journal} {\bibinfo  {journal} {Physical review letters}\
  }\textbf {\bibinfo {volume} {120}},\ \bibinfo {pages} {036002} (\bibinfo
  {year} {2018})}\BibitemShut {NoStop}%
\bibitem [{\citenamefont {Weiler}\ \emph {et~al.}(2018)\citenamefont {Weiler},
  \citenamefont {Geiger}, \citenamefont {Welling}, \citenamefont {Boomsma},\
  and\ \citenamefont {Cohen}}]{weiler20183d}%
  \BibitemOpen
  \bibfield  {author} {\bibinfo {author} {\bibfnamefont {M.}~\bibnamefont
  {Weiler}}, \bibinfo {author} {\bibfnamefont {M.}~\bibnamefont {Geiger}},
  \bibinfo {author} {\bibfnamefont {M.}~\bibnamefont {Welling}}, \bibinfo
  {author} {\bibfnamefont {W.}~\bibnamefont {Boomsma}}, \ and\ \bibinfo
  {author} {\bibfnamefont {T.~S.}\ \bibnamefont {Cohen}},\ }\bibfield  {title}
  {\enquote {\bibinfo {title} {3d steerable cnns: Learning rotationally
  equivariant features in volumetric data},}\ }\href@noop {} {\bibfield
  {journal} {\bibinfo  {journal} {Advances in Neural Information Processing
  Systems}\ }\textbf {\bibinfo {volume} {31}} (\bibinfo {year}
  {2018})}\BibitemShut {NoStop}%
\bibitem [{\citenamefont {Treutler}\ and\ \citenamefont
  {Ahlrichs}(1995)}]{ahrichs1995integration}%
  \BibitemOpen
  \bibfield  {author} {\bibinfo {author} {\bibfnamefont {O.}~\bibnamefont
  {Treutler}}\ and\ \bibinfo {author} {\bibfnamefont {R.}~\bibnamefont
  {Ahlrichs}},\ }\bibfield  {title} {\enquote {\bibinfo {title} {{Efficient
  molecular numerical integration schemes}},}\ }\href {\doibase
  10.1063/1.469408} {\bibfield  {journal} {\bibinfo  {journal} {The Journal of
  Chemical Physics}\ }\textbf {\bibinfo {volume} {102}},\ \bibinfo {pages}
  {346--354} (\bibinfo {year} {1995})},\ \Eprint
  {http://arxiv.org/abs/https://pubs.aip.org/aip/jcp/article-pdf/102/1/346/8103924/346\_1\_online.pdf}
  {https://pubs.aip.org/aip/jcp/article-pdf/102/1/346/8103924/346\_1\_online.pdf}
  \BibitemShut {NoStop}%
\bibitem [{\citenamefont {Becke}(1988)}]{becke1988density}%
  \BibitemOpen
  \bibfield  {author} {\bibinfo {author} {\bibfnamefont {A.~D.}\ \bibnamefont
  {Becke}},\ }\bibfield  {title} {\enquote {\bibinfo {title}
  {Density-functional exchange-energy approximation with correct asymptotic
  behavior},}\ }\href@noop {} {\bibfield  {journal} {\bibinfo  {journal}
  {Physical review A}\ }\textbf {\bibinfo {volume} {38}},\ \bibinfo {pages}
  {3098} (\bibinfo {year} {1988})}\BibitemShut {NoStop}%
\bibitem [{\citenamefont {Lee}, \citenamefont {Yang},\ and\ \citenamefont
  {Parr}(1988)}]{lee1988development}%
  \BibitemOpen
  \bibfield  {author} {\bibinfo {author} {\bibfnamefont {C.}~\bibnamefont
  {Lee}}, \bibinfo {author} {\bibfnamefont {W.}~\bibnamefont {Yang}}, \ and\
  \bibinfo {author} {\bibfnamefont {R.~G.}\ \bibnamefont {Parr}},\ }\bibfield
  {title} {\enquote {\bibinfo {title} {Development of the colle-salvetti
  correlation-energy formula into a functional of the electron density},}\
  }\href@noop {} {\bibfield  {journal} {\bibinfo  {journal} {Physical review
  B}\ }\textbf {\bibinfo {volume} {37}},\ \bibinfo {pages} {785} (\bibinfo
  {year} {1988})}\BibitemShut {NoStop}%
\bibitem [{\citenamefont {Dunning}(1989)}]{dunning1989a}%
  \BibitemOpen
  \bibfield  {author} {\bibinfo {author} {\bibfnamefont {T.~H.}\ \bibnamefont
  {Dunning}},\ }\bibfield  {title} {\enquote {\bibinfo {title} {Gaussian basis
  sets for use in correlated molecular calculations. i. the atoms boron through
  neon and hydrogen},}\ }\href {\doibase 10.1063/1.456153} {\bibfield
  {journal} {\bibinfo  {journal} {J. Chem. Phys.}\ }\textbf {\bibinfo {volume}
  {90}},\ \bibinfo {pages} {1007--1023} (\bibinfo {year} {1989})}\BibitemShut
  {NoStop}%
\bibitem [{\citenamefont {Woon}\ and\ \citenamefont
  {Dunning}(1994)}]{woon1994a}%
  \BibitemOpen
  \bibfield  {author} {\bibinfo {author} {\bibfnamefont {D.~E.}\ \bibnamefont
  {Woon}}\ and\ \bibinfo {author} {\bibfnamefont {T.~H.}\ \bibnamefont
  {Dunning}},\ }\bibfield  {title} {\enquote {\bibinfo {title} {Gaussian basis
  sets for use in correlated molecular calculations. iv. calculation of static
  electrical response properties},}\ }\href {\doibase 10.1063/1.466439}
  {\bibfield  {journal} {\bibinfo  {journal} {J. Chem. Phys.}\ }\textbf
  {\bibinfo {volume} {100}},\ \bibinfo {pages} {2975--2988} (\bibinfo {year}
  {1994})}\BibitemShut {NoStop}%
\bibitem [{\citenamefont {King}\ and\ \citenamefont
  {Handy}(2000)}]{king2000kinetic}%
  \BibitemOpen
  \bibfield  {author} {\bibinfo {author} {\bibfnamefont {R.~A.}\ \bibnamefont
  {King}}\ and\ \bibinfo {author} {\bibfnamefont {N.~C.}\ \bibnamefont
  {Handy}},\ }\bibfield  {title} {\enquote {\bibinfo {title} {Kinetic energy
  functionals from the kohn--sham potential},}\ }\href@noop {} {\bibfield
  {journal} {\bibinfo  {journal} {Physical Chemistry Chemical Physics}\
  }\textbf {\bibinfo {volume} {2}},\ \bibinfo {pages} {5049--5056} (\bibinfo
  {year} {2000})}\BibitemShut {NoStop}%
\bibitem [{\citenamefont {Hohenberg}\ and\ \citenamefont
  {Kohn}(1964)}]{hohenberg_inhomogeneous_1964}%
  \BibitemOpen
  \bibfield  {author} {\bibinfo {author} {\bibfnamefont {P.}~\bibnamefont
  {Hohenberg}}\ and\ \bibinfo {author} {\bibfnamefont {W.}~\bibnamefont
  {Kohn}},\ }\bibfield  {title} {\enquote {\bibinfo {title} {Inhomogeneous
  electron gas},}\ }\href {\doibase 10.1103/PhysRev.136.B864} {\bibfield
  {journal} {\bibinfo  {journal} {Phys. Rev.}\ }\textbf {\bibinfo {volume}
  {136}},\ \bibinfo {pages} {B864--B871} (\bibinfo {year} {1964})}\BibitemShut
  {NoStop}%
\bibitem [{\citenamefont {Sun}\ \emph {et~al.}(2018)\citenamefont {Sun},
  \citenamefont {Berkelbach}, \citenamefont {Blunt}, \citenamefont {Booth},
  \citenamefont {Guo}, \citenamefont {Li}, \citenamefont {Liu}, \citenamefont
  {McClain}, \citenamefont {Sayfutyarova}, \citenamefont {Sharma} \emph
  {et~al.}}]{sun2018pyscf}%
  \BibitemOpen
  \bibfield  {author} {\bibinfo {author} {\bibfnamefont {Q.}~\bibnamefont
  {Sun}}, \bibinfo {author} {\bibfnamefont {T.~C.}\ \bibnamefont {Berkelbach}},
  \bibinfo {author} {\bibfnamefont {N.~S.}\ \bibnamefont {Blunt}}, \bibinfo
  {author} {\bibfnamefont {G.~H.}\ \bibnamefont {Booth}}, \bibinfo {author}
  {\bibfnamefont {S.}~\bibnamefont {Guo}}, \bibinfo {author} {\bibfnamefont
  {Z.}~\bibnamefont {Li}}, \bibinfo {author} {\bibfnamefont {J.}~\bibnamefont
  {Liu}}, \bibinfo {author} {\bibfnamefont {J.~D.}\ \bibnamefont {McClain}},
  \bibinfo {author} {\bibfnamefont {E.~R.}\ \bibnamefont {Sayfutyarova}},
  \bibinfo {author} {\bibfnamefont {S.}~\bibnamefont {Sharma}},  \emph
  {et~al.},\ }\bibfield  {title} {\enquote {\bibinfo {title} {Pyscf: the
  python-based simulations of chemistry framework},}\ }\href@noop {} {\bibfield
   {journal} {\bibinfo  {journal} {Wiley Interdisciplinary Reviews:
  Computational Molecular Science}\ }\textbf {\bibinfo {volume} {8}},\ \bibinfo
  {pages} {e1340} (\bibinfo {year} {2018})}\BibitemShut {NoStop}%
\bibitem [{\citenamefont {Sun}\ \emph {et~al.}(2020)\citenamefont {Sun},
  \citenamefont {Zhang}, \citenamefont {Banerjee}, \citenamefont {Bao},
  \citenamefont {Barbry}, \citenamefont {Blunt}, \citenamefont {Bogdanov},
  \citenamefont {Booth}, \citenamefont {Chen}, \citenamefont {Cui} \emph
  {et~al.}}]{sun2020recent}%
  \BibitemOpen
  \bibfield  {author} {\bibinfo {author} {\bibfnamefont {Q.}~\bibnamefont
  {Sun}}, \bibinfo {author} {\bibfnamefont {X.}~\bibnamefont {Zhang}}, \bibinfo
  {author} {\bibfnamefont {S.}~\bibnamefont {Banerjee}}, \bibinfo {author}
  {\bibfnamefont {P.}~\bibnamefont {Bao}}, \bibinfo {author} {\bibfnamefont
  {M.}~\bibnamefont {Barbry}}, \bibinfo {author} {\bibfnamefont {N.~S.}\
  \bibnamefont {Blunt}}, \bibinfo {author} {\bibfnamefont {N.~A.}\ \bibnamefont
  {Bogdanov}}, \bibinfo {author} {\bibfnamefont {G.~H.}\ \bibnamefont {Booth}},
  \bibinfo {author} {\bibfnamefont {J.}~\bibnamefont {Chen}}, \bibinfo {author}
  {\bibfnamefont {Z.-H.}\ \bibnamefont {Cui}},  \emph {et~al.},\ }\bibfield
  {title} {\enquote {\bibinfo {title} {Recent developments in the pyscf program
  package},}\ }\href@noop {} {\bibfield  {journal} {\bibinfo  {journal} {The
  Journal of chemical physics}\ }\textbf {\bibinfo {volume} {153}},\ \bibinfo
  {pages} {024109} (\bibinfo {year} {2020})}\BibitemShut {NoStop}%
\bibitem [{\citenamefont {Chan}, \citenamefont {Cohen},\ and\ \citenamefont
  {Handy}(2001)}]{chan_thomasfermidiracvon_2001}%
  \BibitemOpen
  \bibfield  {author} {\bibinfo {author} {\bibfnamefont {G.~K.-L.}\
  \bibnamefont {Chan}}, \bibinfo {author} {\bibfnamefont {A.~J.}\ \bibnamefont
  {Cohen}}, \ and\ \bibinfo {author} {\bibfnamefont {N.~C.}\ \bibnamefont
  {Handy}},\ }\bibfield  {title} {\enquote {\bibinfo {title}
  {{Thomas–Fermi–Dirac–von Weizsäcker} models in finite systems},}\
  }\href {\doibase 10.1063/1.1321308} {\bibfield  {journal} {\bibinfo
  {journal} {J. Chem. Phys.}\ }\textbf {\bibinfo {volume} {114}},\ \bibinfo
  {pages} {631} (\bibinfo {year} {2001})}\BibitemShut {NoStop}%
\bibitem [{\citenamefont {Ryley}\ \emph {et~al.}(2021)\citenamefont {Ryley},
  \citenamefont {Withnall}, \citenamefont {Irons}, \citenamefont {Helgaker},\
  and\ \citenamefont {Teale}}]{ryley_robust_2021}%
  \BibitemOpen
  \bibfield  {author} {\bibinfo {author} {\bibfnamefont {M.~S.}\ \bibnamefont
  {Ryley}}, \bibinfo {author} {\bibfnamefont {M.}~\bibnamefont {Withnall}},
  \bibinfo {author} {\bibfnamefont {T.~J.~P.}\ \bibnamefont {Irons}}, \bibinfo
  {author} {\bibfnamefont {T.}~\bibnamefont {Helgaker}}, \ and\ \bibinfo
  {author} {\bibfnamefont {A.~M.}\ \bibnamefont {Teale}},\ }\bibfield  {title}
  {\enquote {\bibinfo {title} {Robust all-electron optimization in orbital-free
  density-functional theory using the trust-region image method},}\ }\href
  {\doibase 10.1021/acs.jpca.0c09502} {\bibfield  {journal} {\bibinfo
  {journal} {J. Phys. Chem. A}\ }\textbf {\bibinfo {volume} {125}},\ \bibinfo
  {pages} {459--475} (\bibinfo {year} {2021})},\ \bibinfo {note} {publisher:
  American Chemical Society}\BibitemShut {NoStop}%
\bibitem [{\citenamefont {Kraft}(1988)}]{kraft1988software}%
  \BibitemOpen
  \bibfield  {author} {\bibinfo {author} {\bibfnamefont {D.}~\bibnamefont
  {Kraft}},\ }\href {https://books.google.de/books?id=4rKaGwAACAAJ} {\emph
  {\bibinfo {title} {A Software Package for Sequential Quadratic
  Programming}}},\ Deutsche Forschungs- und Versuchsanstalt f{\"u}r Luft- und
  Raumfahrt K{\"o}ln: Forschungsbericht\ (\bibinfo  {publisher} {Wiss.
  Berichtswesen d. DFVLR},\ \bibinfo {year} {1988})\BibitemShut {NoStop}%
\bibitem [{\citenamefont {Virtanen}\ \emph {et~al.}(2020)\citenamefont
  {Virtanen}, \citenamefont {Gommers}, \citenamefont {Oliphant}, \citenamefont
  {Haberland}, \citenamefont {Reddy}, \citenamefont {Cournapeau}, \citenamefont
  {Burovski}, \citenamefont {Peterson}, \citenamefont {Weckesser},
  \citenamefont {Bright}, \citenamefont {{van der Walt}}, \citenamefont
  {Brett}, \citenamefont {Wilson}, \citenamefont {Millman}, \citenamefont
  {Mayorov}, \citenamefont {Nelson}, \citenamefont {Jones}, \citenamefont
  {Kern}, \citenamefont {Larson}, \citenamefont {Carey}, \citenamefont {Polat},
  \citenamefont {Feng}, \citenamefont {Moore}, \citenamefont {{VanderPlas}},
  \citenamefont {Laxalde}, \citenamefont {Perktold}, \citenamefont {Cimrman},
  \citenamefont {Henriksen}, \citenamefont {Quintero}, \citenamefont {Harris},
  \citenamefont {Archibald}, \citenamefont {Ribeiro}, \citenamefont
  {Pedregosa}, \citenamefont {{van Mulbregt}},\ and\ \citenamefont {{SciPy 1.0
  Contributors}}}]{2020SciPy-NMeth}%
  \BibitemOpen
  \bibfield  {author} {\bibinfo {author} {\bibfnamefont {P.}~\bibnamefont
  {Virtanen}}, \bibinfo {author} {\bibfnamefont {R.}~\bibnamefont {Gommers}},
  \bibinfo {author} {\bibfnamefont {T.~E.}\ \bibnamefont {Oliphant}}, \bibinfo
  {author} {\bibfnamefont {M.}~\bibnamefont {Haberland}}, \bibinfo {author}
  {\bibfnamefont {T.}~\bibnamefont {Reddy}}, \bibinfo {author} {\bibfnamefont
  {D.}~\bibnamefont {Cournapeau}}, \bibinfo {author} {\bibfnamefont
  {E.}~\bibnamefont {Burovski}}, \bibinfo {author} {\bibfnamefont
  {P.}~\bibnamefont {Peterson}}, \bibinfo {author} {\bibfnamefont
  {W.}~\bibnamefont {Weckesser}}, \bibinfo {author} {\bibfnamefont
  {J.}~\bibnamefont {Bright}}, \bibinfo {author} {\bibfnamefont {S.~J.}\
  \bibnamefont {{van der Walt}}}, \bibinfo {author} {\bibfnamefont
  {M.}~\bibnamefont {Brett}}, \bibinfo {author} {\bibfnamefont
  {J.}~\bibnamefont {Wilson}}, \bibinfo {author} {\bibfnamefont {K.~J.}\
  \bibnamefont {Millman}}, \bibinfo {author} {\bibfnamefont {N.}~\bibnamefont
  {Mayorov}}, \bibinfo {author} {\bibfnamefont {A.~R.~J.}\ \bibnamefont
  {Nelson}}, \bibinfo {author} {\bibfnamefont {E.}~\bibnamefont {Jones}},
  \bibinfo {author} {\bibfnamefont {R.}~\bibnamefont {Kern}}, \bibinfo {author}
  {\bibfnamefont {E.}~\bibnamefont {Larson}}, \bibinfo {author} {\bibfnamefont
  {C.~J.}\ \bibnamefont {Carey}}, \bibinfo {author} {\bibfnamefont
  {{\.I}.}~\bibnamefont {Polat}}, \bibinfo {author} {\bibfnamefont
  {Y.}~\bibnamefont {Feng}}, \bibinfo {author} {\bibfnamefont {E.~W.}\
  \bibnamefont {Moore}}, \bibinfo {author} {\bibfnamefont {J.}~\bibnamefont
  {{VanderPlas}}}, \bibinfo {author} {\bibfnamefont {D.}~\bibnamefont
  {Laxalde}}, \bibinfo {author} {\bibfnamefont {J.}~\bibnamefont {Perktold}},
  \bibinfo {author} {\bibfnamefont {R.}~\bibnamefont {Cimrman}}, \bibinfo
  {author} {\bibfnamefont {I.}~\bibnamefont {Henriksen}}, \bibinfo {author}
  {\bibfnamefont {E.~A.}\ \bibnamefont {Quintero}}, \bibinfo {author}
  {\bibfnamefont {C.~R.}\ \bibnamefont {Harris}}, \bibinfo {author}
  {\bibfnamefont {A.~M.}\ \bibnamefont {Archibald}}, \bibinfo {author}
  {\bibfnamefont {A.~H.}\ \bibnamefont {Ribeiro}}, \bibinfo {author}
  {\bibfnamefont {F.}~\bibnamefont {Pedregosa}}, \bibinfo {author}
  {\bibfnamefont {P.}~\bibnamefont {{van Mulbregt}}}, \ and\ \bibinfo {author}
  {\bibnamefont {{SciPy 1.0 Contributors}}},\ }\bibfield  {title} {\enquote
  {\bibinfo {title} {{{SciPy} 1.0: Fundamental Algorithms for Scientific
  Computing in Python}},}\ }\href {\doibase 10.1038/s41592-019-0686-2}
  {\bibfield  {journal} {\bibinfo  {journal} {Nature Methods}\ }\textbf
  {\bibinfo {volume} {17}},\ \bibinfo {pages} {261--272} (\bibinfo {year}
  {2020})}\BibitemShut {NoStop}%
\bibitem [{\citenamefont {Kingma}\ and\ \citenamefont
  {Ba}(2014)}]{kingma2014adam}%
  \BibitemOpen
  \bibfield  {author} {\bibinfo {author} {\bibfnamefont {D.~P.}\ \bibnamefont
  {Kingma}}\ and\ \bibinfo {author} {\bibfnamefont {J.}~\bibnamefont {Ba}},\
  }\bibfield  {title} {\enquote {\bibinfo {title} {Adam: A method for
  stochastic optimization},}\ }\href@noop {} {\bibfield  {journal} {\bibinfo
  {journal} {arXiv preprint arXiv:1412.6980}\ } (\bibinfo {year}
  {2014})}\BibitemShut {NoStop}%
\bibitem [{\citenamefont {Canc{\`e}s}, \citenamefont {Maday},\ and\
  \citenamefont {Stamm}(2013)}]{cances2013domain}%
  \BibitemOpen
  \bibfield  {author} {\bibinfo {author} {\bibfnamefont {E.}~\bibnamefont
  {Canc{\`e}s}}, \bibinfo {author} {\bibfnamefont {Y.}~\bibnamefont {Maday}}, \
  and\ \bibinfo {author} {\bibfnamefont {B.}~\bibnamefont {Stamm}},\ }\bibfield
   {title} {\enquote {\bibinfo {title} {Domain decomposition for implicit
  solvation models},}\ }\href@noop {} {\bibfield  {journal} {\bibinfo
  {journal} {The Journal of chemical physics}\ }\textbf {\bibinfo {volume}
  {139}},\ \bibinfo {pages} {054111} (\bibinfo {year} {2013})}\BibitemShut
  {NoStop}%
\bibitem [{\citenamefont {Lipparini}\ \emph {et~al.}(2013)\citenamefont
  {Lipparini}, \citenamefont {Stamm}, \citenamefont {Cances}, \citenamefont
  {Maday},\ and\ \citenamefont {Mennucci}}]{lipparini2013fast}%
  \BibitemOpen
  \bibfield  {author} {\bibinfo {author} {\bibfnamefont {F.}~\bibnamefont
  {Lipparini}}, \bibinfo {author} {\bibfnamefont {B.}~\bibnamefont {Stamm}},
  \bibinfo {author} {\bibfnamefont {E.}~\bibnamefont {Cances}}, \bibinfo
  {author} {\bibfnamefont {Y.}~\bibnamefont {Maday}}, \ and\ \bibinfo {author}
  {\bibfnamefont {B.}~\bibnamefont {Mennucci}},\ }\bibfield  {title} {\enquote
  {\bibinfo {title} {Fast domain decomposition algorithm for continuum
  solvation models: Energy and first derivatives},}\ }\href@noop {} {\bibfield
  {journal} {\bibinfo  {journal} {Journal of chemical theory and computation}\
  }\textbf {\bibinfo {volume} {9}},\ \bibinfo {pages} {3637--3648} (\bibinfo
  {year} {2013})}\BibitemShut {NoStop}%
\bibitem [{\citenamefont {Lipparini}\ \emph {et~al.}(2014)\citenamefont
  {Lipparini}, \citenamefont {Scalmani}, \citenamefont {Lagard{\`e}re},
  \citenamefont {Stamm}, \citenamefont {Canc{\`e}s}, \citenamefont {Maday},
  \citenamefont {Piquemal}, \citenamefont {Frisch},\ and\ \citenamefont
  {Mennucci}}]{lipparini2014quantum}%
  \BibitemOpen
  \bibfield  {author} {\bibinfo {author} {\bibfnamefont {F.}~\bibnamefont
  {Lipparini}}, \bibinfo {author} {\bibfnamefont {G.}~\bibnamefont {Scalmani}},
  \bibinfo {author} {\bibfnamefont {L.}~\bibnamefont {Lagard{\`e}re}}, \bibinfo
  {author} {\bibfnamefont {B.}~\bibnamefont {Stamm}}, \bibinfo {author}
  {\bibfnamefont {E.}~\bibnamefont {Canc{\`e}s}}, \bibinfo {author}
  {\bibfnamefont {Y.}~\bibnamefont {Maday}}, \bibinfo {author} {\bibfnamefont
  {J.-P.}\ \bibnamefont {Piquemal}}, \bibinfo {author} {\bibfnamefont {M.~J.}\
  \bibnamefont {Frisch}}, \ and\ \bibinfo {author} {\bibfnamefont
  {B.}~\bibnamefont {Mennucci}},\ }\bibfield  {title} {\enquote {\bibinfo
  {title} {Quantum, classical, and hybrid qm/mm calculations in solution:
  General implementation of the ddcosmo linear scaling strategy},}\ }\href@noop
  {} {\bibfield  {journal} {\bibinfo  {journal} {The Journal of chemical
  physics}\ }\textbf {\bibinfo {volume} {141}},\ \bibinfo {pages} {184108}
  (\bibinfo {year} {2014})}\BibitemShut {NoStop}%
\end{thebibliography}%

\section{Appendix}

\subsection{Data Generation}

As mentioned in section \ref{sec:DataGeneration}, to generate the training data for our model we slightly perturbed the external potential of our molecules to sample a diverse set of densities as solution of Kohn-Sham-DFT and thereby calculate our targets.
To achieve this we used the pyscf package as code base and implemented our own Restricted Kohn Sham class which takes an additional Matrix and adds it to the external potential matrix in the Hamiltonian of the SCF procedure.

For the sampling of those perturbation matrices the following approach was adopted:

\begin{enumerate}
    \item (Relative) entries of the perturbation matrix are drawn from some random distribution and ensure a symmetric matrix
    \item the norm of the perturbation matrix is drawn from some distribution and the matrix is normalized accordingly
\end{enumerate}

\begin{figure*}[p]
    \begin{subfigure}[c]{.8\columnwidth}
    \includegraphicscorrection[width=\textwidth]{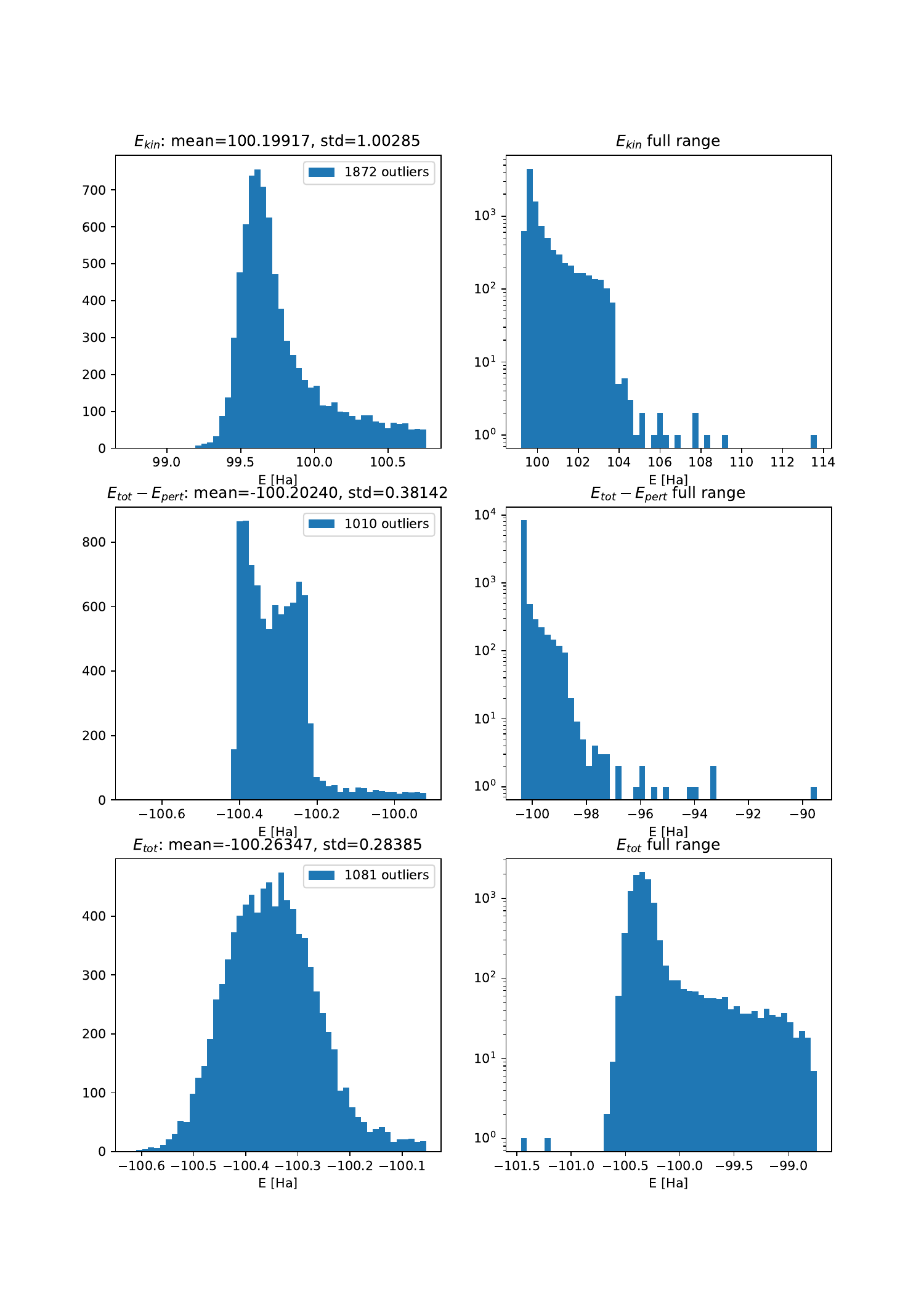}
    \subcaption{\ce{HF}}
    \end{subfigure}
    \begin{subfigure}[c]{.8\columnwidth}
    \includegraphicscorrection[width=\textwidth]{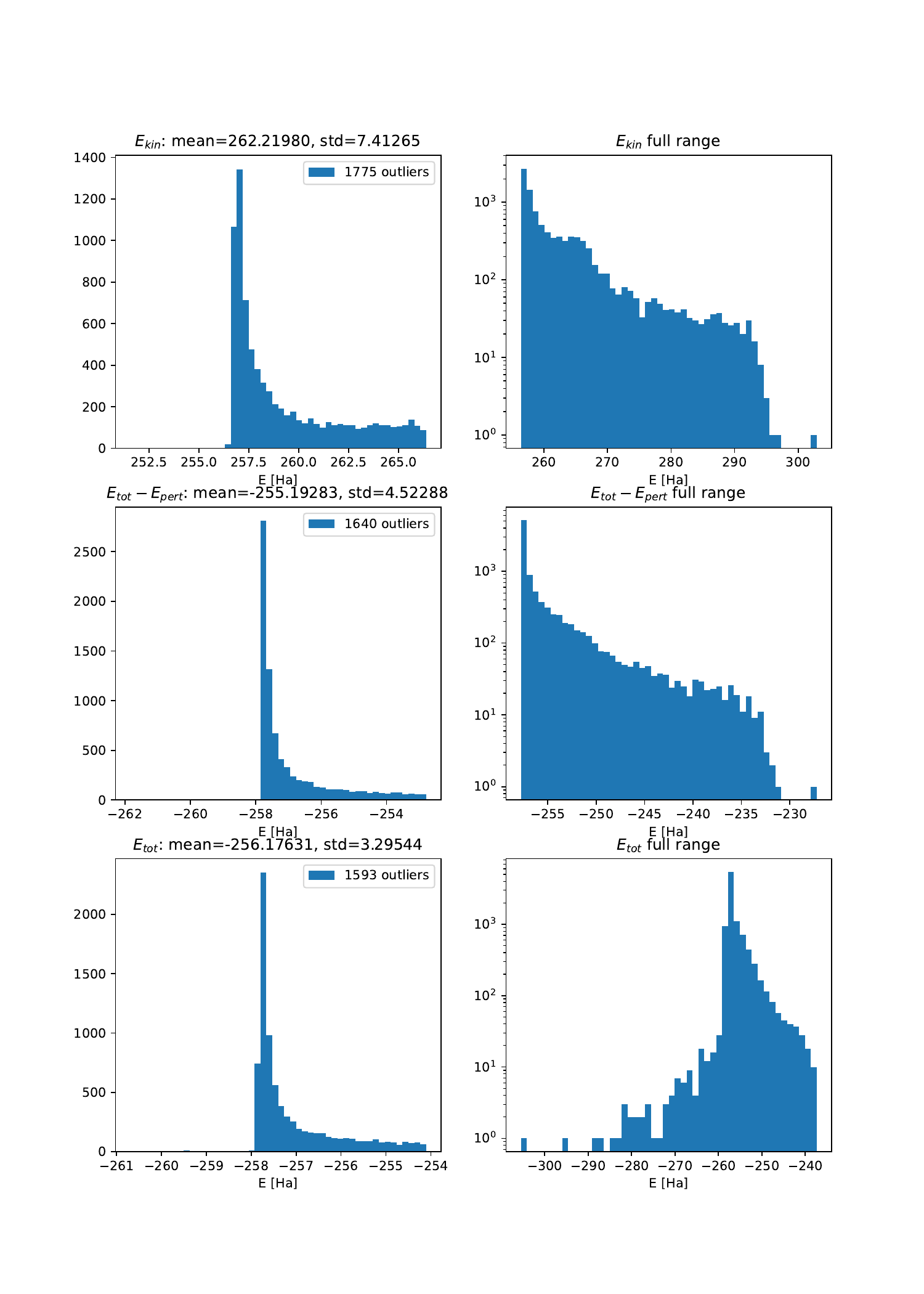}
    \subcaption{\ce{Ne2}}
    \end{subfigure}
    
    \begin{subfigure}[c]{.8\columnwidth}
    \includegraphicscorrection[width=\textwidth]{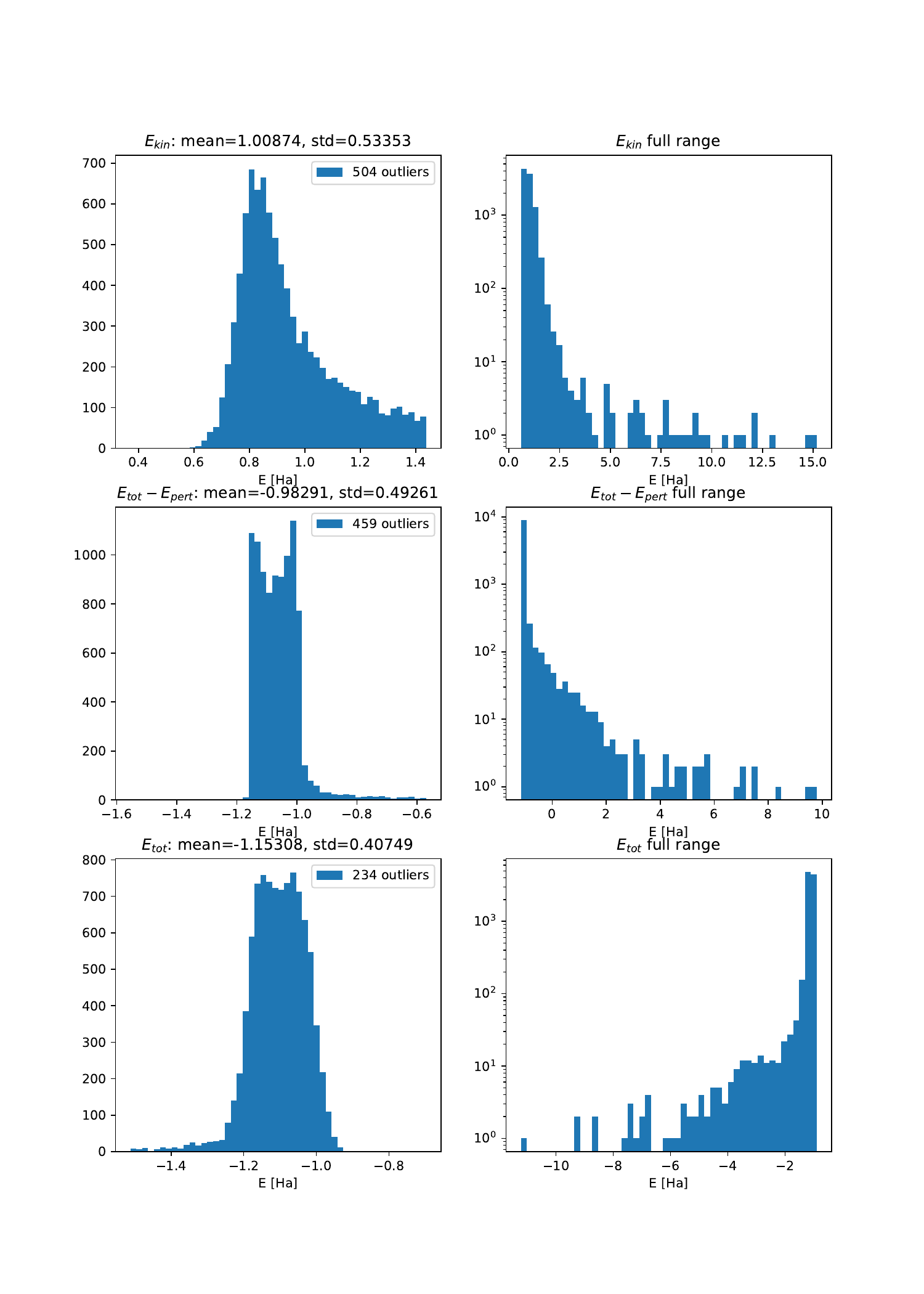}
    \subcaption{\ce{H2}}
    \end{subfigure}
    \begin{subfigure}[c]{.8\columnwidth}
    \includegraphicscorrection[width=\textwidth]{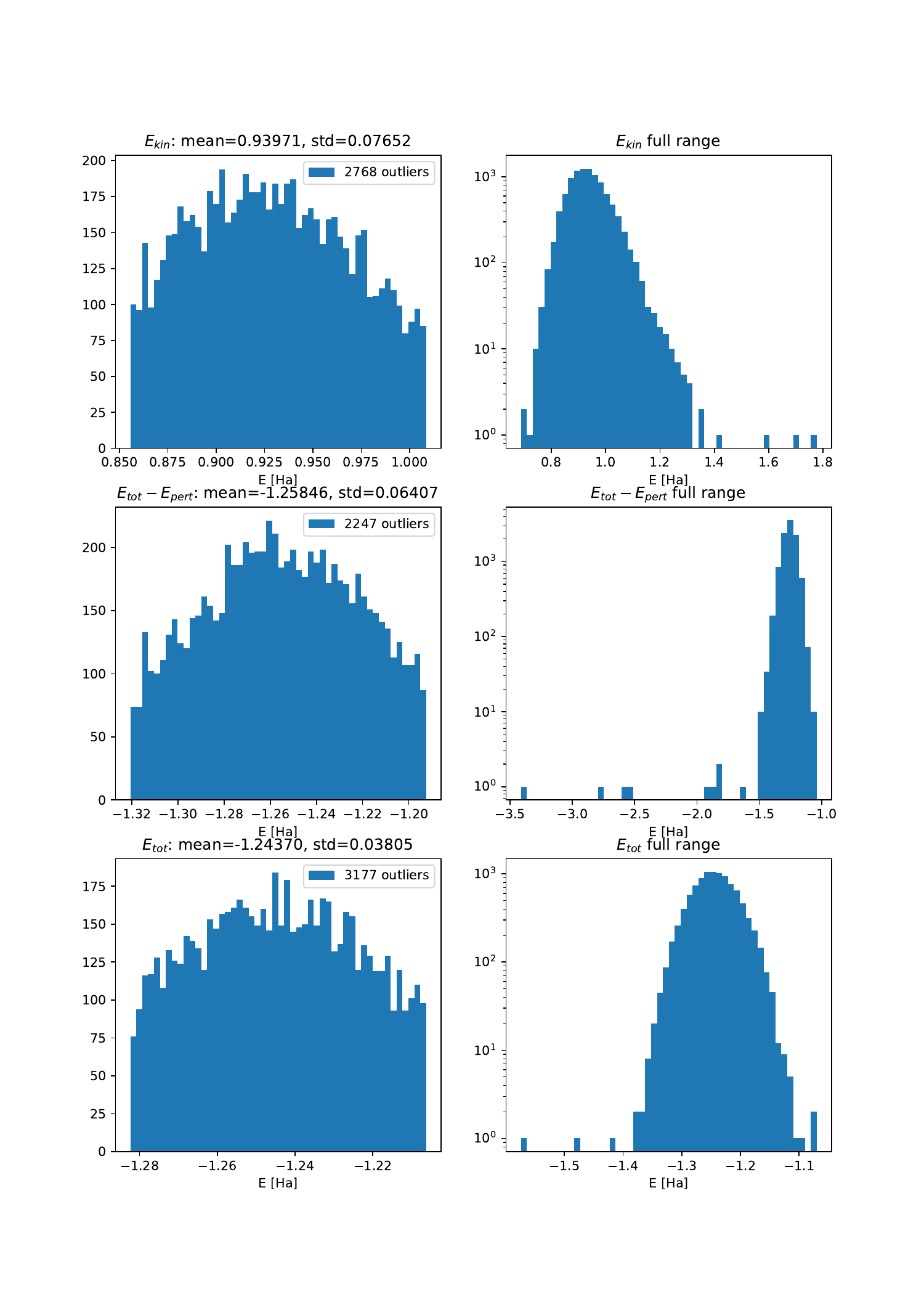}
    \subcaption{\ce{H3+}}
    \end{subfigure}
    \caption{Distribution of the kinetic energy $E_\text{kin}$, 
    the total energy without the contribution from our perturbation to the external potential $E_\text{tot}-E_\text{pert}$ and the total energy.}
    \label{fig:e_dist_1}
\end{figure*}

\begin{figure*}[p]
    \begin{subfigure}[c]{.8\columnwidth}
    \includegraphicscorrection[width=\textwidth]{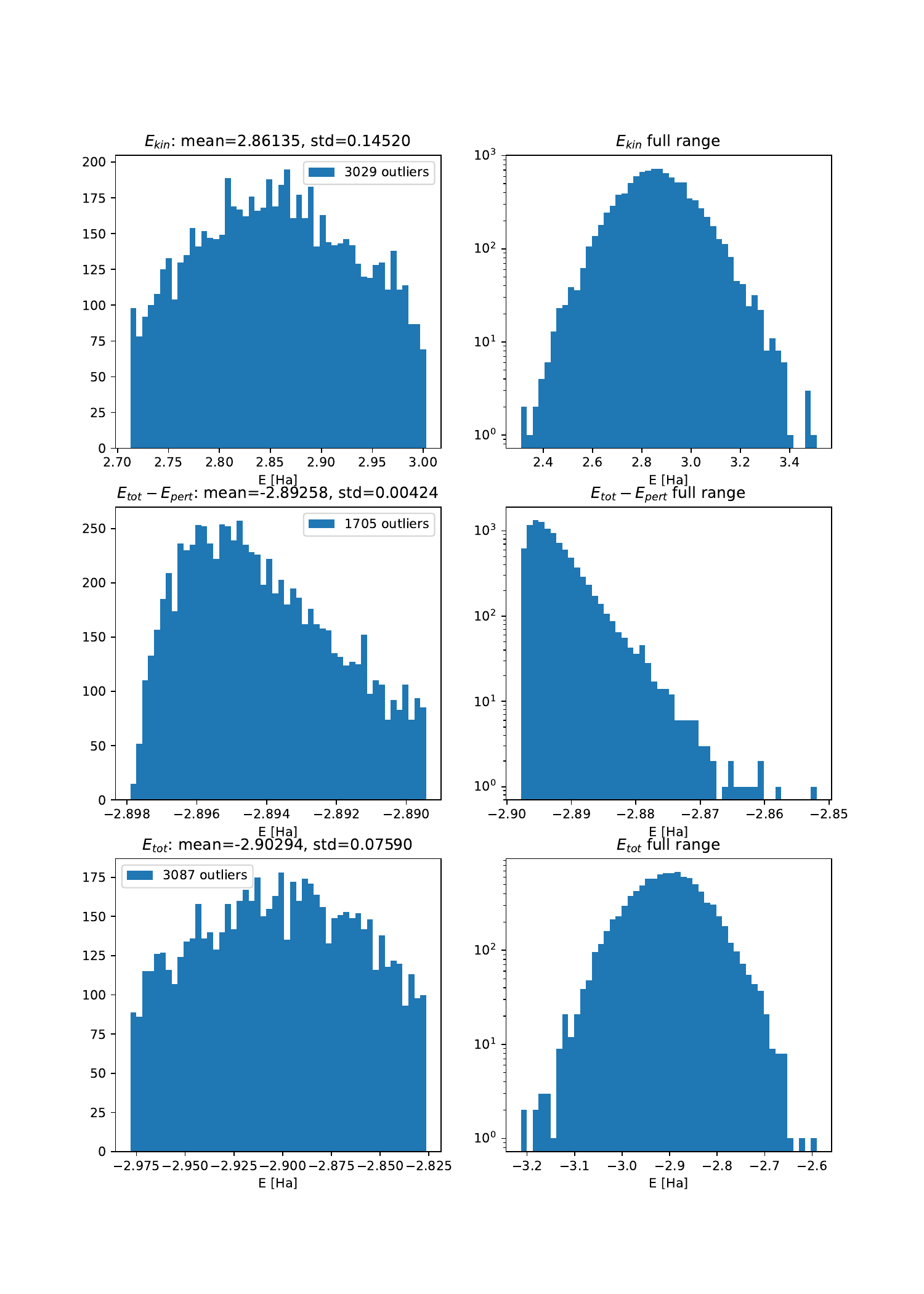}
    \subcaption{\ce{He}}
    \end{subfigure}
    \begin{subfigure}[c]{.8\columnwidth}
    \includegraphicscorrection[width=\textwidth]{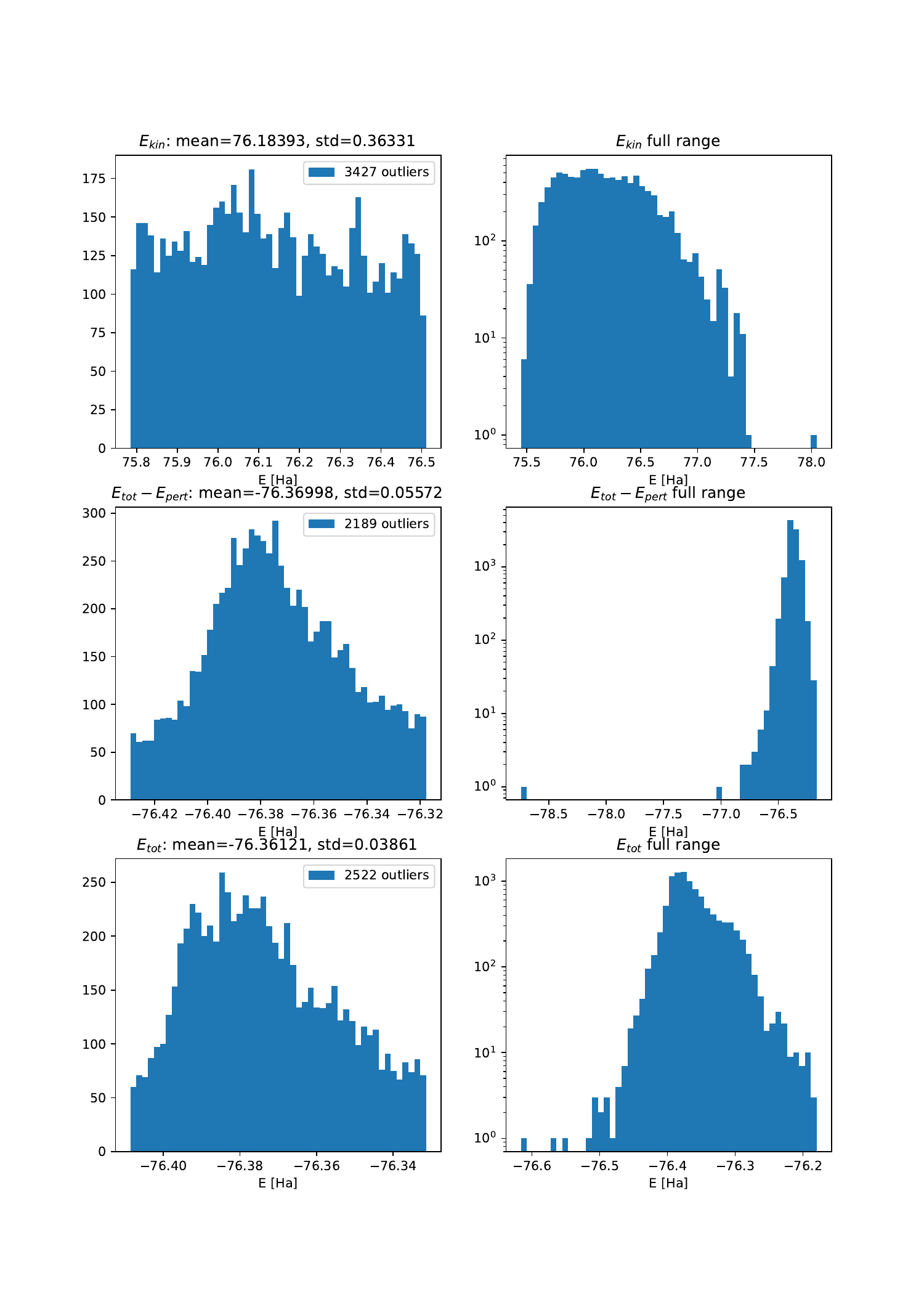}
    \subcaption{\ce{H2O}}
    \end{subfigure}
    
    \caption{ Distribution of the kinetic energy $E_\text{kin}$, 
    the total energy without the contribution from our perturbation to the external potential $E_\text{tot}-E_\text{pert}$ and the total energy.}
    \label{fig:e_dist_2}
\end{figure*}

The distributions are chosen to ensure that a majority of data points are somewhat close to the ground state.
The reasoning being that for accurate convergence and ground state values the machine learning model should make precise predictions in this part of density space while far away from the solution a rough estimate of the kinetic energy and potential should be enough to guide the OF-DFT solver in the correct direction.
Details regarding those distributions in the different basis sets are given in table \ref{tab:datasets_details}. All data sets have been calculated using at the BLYP/cc-pVDZ level of theory. 
The distribution of the kinetic and total energies in those data sets are shown in figures \correction{\ref{fig:e_dist_1} and \ref{fig:e_dist_2}.}

\begin{table}[h]
    \caption{Data sets used for training. The norm of the perturbation matrix is sampled from a normal distribution with a minimum norm of 0.005 and the mean and standard deviation given in the table. The relative entries of these matrices are sampled from a normal distribution with mean and standard deviation given in the table.}
    \label{tab:datasets_details}
    \centering
    \begin{tabular}{c|cc|cc}
    \multirow{2}{*}{Data set}    & \multicolumn{2}{c|}{Norm} & \multicolumn{2}{c}{Matrix elements} \\
                                & $\mu$ & $\sigma$ & $\mu$ & $\sigma$ \\
    \ce{HF}     & 0.25 & 0.05 & 0.0 & 0.2 \\ %
    \ce{Ne2}    & 0.25 & 0.05 & 0.0 & 0.2 \\ %
    \ce{H2}     & 0.25 & 0.05 & 0.0 & 0.2 \\ %
    \ce{H3+}    & 0.25 & 0.05 & 0.0 & 0.2 \\
    \ce{He}     & 0.25 & 0.05 & 0.0 & 0.2 \\ %
    \ce{H2O}    & 0.00 & 0.10 & 0.0 & 0.2 \\
    \end{tabular}
\end{table}

\subsection{OF-DFT implementation}
Our OF-DFT implementation is based on the pyscf package, which we use to compute all the required integrals.
Density fitting as implemented in the pyscf package is used for the calculation of the Coulomb matrix.
\correction{

In sporadic cases, we find the optimization gets stuck in a bad local minimum not corresponding to the ground state but to a state with a zero crossing in the single ``orbital" $\phi$ (see eq. \ref{eq:of-dft-ansatz}). While the entire procedure is deterministic in principle, in practice it is not due to numerical noise for instance in certain layers of our machine learned functionals (For example, }\verb|torch.scatter| \correction{is known to have this property, see the pytorch documentation). Hence we simply re-run the density optimization when we detect such zero crossings to find the correct minimum.}

\begin{figure*}[h]
    \centering
    \includegraphicscorrection[width=0.7\textwidth]{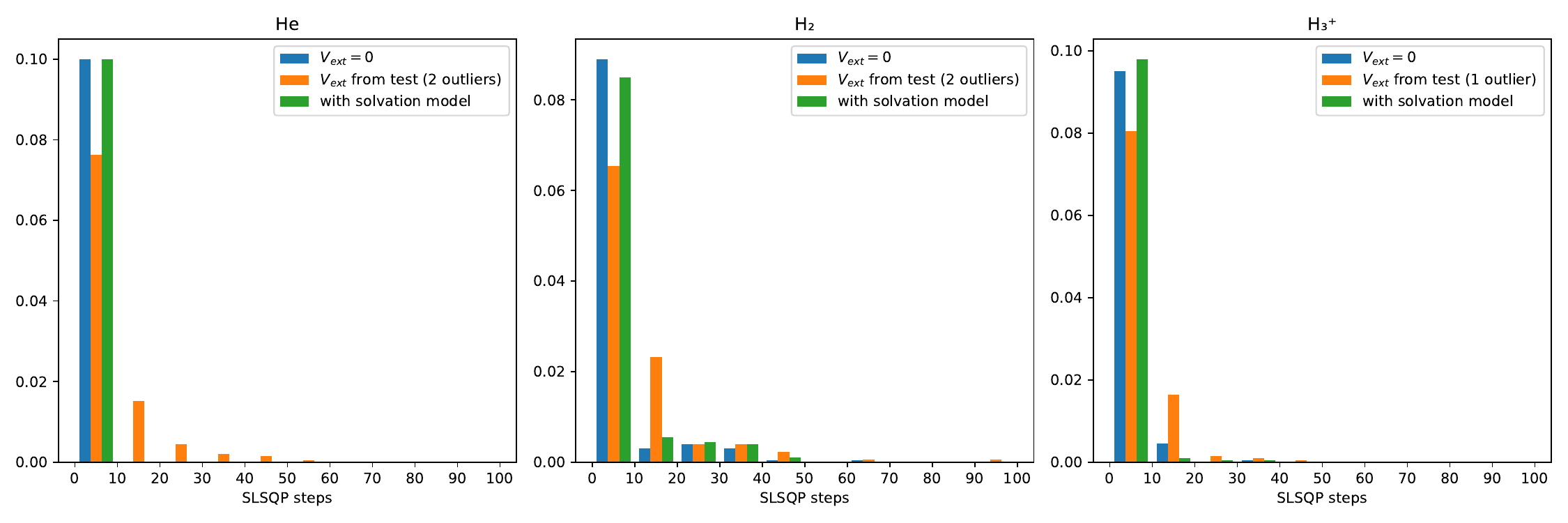}
    \caption{ Distribution of steps until convergence for the three two-electron systems and different density optimization modes.}
    \label{fig:steps_histogram}
\end{figure*}

\subsection{Correspondence between KS and OF Ansatz for two electrons}\label{sec:2e-correspondence}

\setcounter{equation}{0}
\makeatletter

\renewcommand{\theequation}{C\arabic{equation}}

Recall the equation for the electron density in terms of the coefficients $c_{\nu}$ in our OF approach:
\begin{equation}\label{eq:of-dft-ansatz-appendix}
    \rho(\mathbf{r}) = \Big(\sum_{\nu} c_{\nu} \chi_{\nu}(\mathbf{r}) \Big)^2 \quad .
\end{equation}
In KS-DFT one can write the electron density in the basis of atomic basis functions using the molecular orbital coefficients $m_{i\nu}$:
\begin{equation}\label{eq:ks-dft-ansatz-appendix}
    \rho(\mathbf{r}) = \sum_i\Big(\sum_{\nu} m_{i\nu} \chi_{\nu}(\mathbf{r}) \Big)^2 \quad .
\end{equation}
For two electron systems, there is only one molecular orbital, hence the first sum disappears. Thus, the two expressions for the density are equal for $m_{1\nu}=c_{\nu}$.

\subsection{Atomic Contributions}\label{sec:atomic-contributions}
The atomic contributions have been calculated using either restricted or restricted open-shell KS-DFT at the BLYP/cc-pVDZ level of theory.
An ``atomic'' initial guess was used, as implemented in the pyscf package.
The convergence tolerance was set to $10^{-4}$ \correction{hartree} and a grid level of 2 was used.
Symmetry was employed to remove directional bias.
The usage of symmetry ensures separation w.r.t.~angular momentum. 
This allows the following procedure for spherical symmetrization of  p type orbitals:
First the MO coefficients and energies are averaged weighted by their occupation. 
Next the electrons in p-orbitals are evenly distributed over all three p-orbitals.
As this procedure has only been implemented for p type orbitals only elements up to Neon can be used.
\correction{

Using these atomic contributions should not introduce any systematic error as we train our models
to predict the correct difference between the full quantity minus the atomic contribution 
and can thus be used to predict the correct total quantity.}

\subsection{Model Hyperparameters}
We use identical hyperparameters for our model for the kinetic energy and for the kinetic potential. 
Our models employ $L=5$ atom-atom interaction layers.
We choose the number of features per order $l$ such that approximately the same number of floats are used for each $l$. 
For the encoder and decoder, we use features of type (40, 14, 8, 6, 4) (i.e.~40 scalars, 14 vectors, 8 $l=2$ tensors and so on), for all layers in-between features of type (101, 34, 20, 14, 11).
We use a radial basis consisting of 32 functions for the encoder and decoder, and 16 functions for the atom-atom interaction layers.

\end{document}